\documentclass[sigconf,natbib]{acmart}
\acmSubmissionID{1436}
\AtBeginDocument{%
  \providecommand\BibTeX{{%
    \normalfont B\kern-0.5em{\scshape i\kern-0.25em b}\kern-0.8em\TeX}}}

\copyrightyear{2023}
\acmYear{2023}
\setcopyright{acmlicensed}\acmConference[KDD '23]{Proceedings of the 29th ACM SIGKDD Conference on Knowledge Discovery and Data Mining}{August 6--10, 2023}{Long Beach, CA, USA}
\acmBooktitle{Proceedings of the 29th ACM SIGKDD Conference on Knowledge Discovery and Data Mining (KDD '23), August 6--10, 2023, Long Beach, CA, USA}
\acmPrice{15.00}
\acmDOI{10.1145/3580305.3599450} \acmISBN{979-8-4007-0103-0/23/08}

\usepackage{multirow}
\usepackage{multicol}
\usepackage{subfigure}
\usepackage{booktabs}
\usepackage{mathrsfs}
\usepackage{balance}
\usepackage{makecell}
\usepackage{algorithm}
\usepackage{algorithmic}
\usepackage{color, xcolor}

\begin{document}

\title{On Manipulating Signals of User-Item Graph: A Jacobi Polynomial-based Graph Collaborative Filtering}

%% Jacobi Polynomial; 
\author{Jiayan Guo}
\authornote{Work done during the internship at Microsoft.}
\email{guojiayan@pku.edu.cn}
\affiliation{%
  \institution{School of Intelligence Science and Technology, Peking University}
  \streetaddress{P.O. Box 1212}
  \city{Beijing}
  \country{China}
  \postcode{43017-6221}
}

\author{Lun Du}
\authornote{corresponding author}
\email{lun.du@microsoft.com}
\affiliation{%
  \institution{Microsoft Research Asia}
  \streetaddress{1 Th{\o}rv{\"a}ld Circle}
  \city{Beijing}
  \country{China}}

\author{Xu Chen}
\email{xu.chen@microsoft.com}
\affiliation{%
  \institution{Microsoft Research Asia}
  \city{Beijing}
  \country{China}
}

\author{Xiaojun Ma}
\email{xiaojunma@microsoft.com}
\affiliation{%
 \institution{Microsoft Research Asia}
 \streetaddress{Rono-Hills}
 \city{Beijing}
 \country{China}}

\author{Qiang Fu}
\email{qifu@microsoft.com}
\affiliation{%
  \institution{Microsoft Research Asia}
  \streetaddress{30 Shuangqing Rd}
  \city{Beijing}
  \country{China}}

\author{Shi Han}
\email{shihan@microsoft.com}
\affiliation{%
  \institution{Microsoft Research Asia}
  \streetaddress{8600 Datapoint Drive}
  \city{Beijing}
  % \state{Texas}
  \country{China}}

\author{Dongmei Zhang}
\email{dongmeiz@microsoft.com}
\affiliation{%
  \institution{Microsoft Research Asia}
  \streetaddress{1 Th{\o}rv{\"a}ld Circle}
  \city{Beijing}
  \country{China}}

\author{Yan Zhang}
\email{zhyzhy001@pku.edu.cn}
\affiliation{%
  \institution{School of Intelligence Science and Technology, Peking University}
  \city{Beijing}
  \country{China}}

\renewcommand{\shortauthors}{Jiayan, et al.}


\begin{CCSXML}
<ccs2012>
<concept>
<concept_id>10002951.10003317.10003347.10003350</concept_id>
<concept_desc>Information systems~Recommender systems</concept_desc>
<concept_significance>500</concept_significance>
</concept>
 <concept>
  <concept_id>10010520.10010553.10010562</concept_id>
  <concept_desc>Computer systems organization~Embedded systems</concept_desc>
  <concept_significance>500</concept_significance>
 </concept>
 <concept>
  <concept_id>10010520.10010575.10010755</concept_id>
  <concept_desc>Computer systems organization~Redundancy</concept_desc>
  <concept_significance>300</concept_significance>
 </concept>
 <concept>
  <concept_id>10010520.10010553.10010554</concept_id>
  <concept_desc>Computer systems organization~Robotics</concept_desc>
  <concept_significance>100</concept_significance>
 </concept>
 <concept>
  <concept_id>10003033.10003083.10003095</concept_id>
  <concept_desc>Networks~Network reliability</concept_desc>
  <concept_significance>100</concept_significance>
 </concept>
</ccs2012>
\end{CCSXML}

\ccsdesc[500]{Information systems~Recommender systems}

\keywords{Recommender System, Graph Collaborative Filtering}


\begin{abstract}
  Collaborative filtering~(CF) is an important research direction in recommender systems that aims to make recommendations given the information on user-item interactions. Graph CF has attracted more and more attention in recent years due to its effectiveness in leveraging high-order information in the user-item bipartite graph for better recommendations. Specifically, recent studies show the success of graph neural networks~(GNN) for CF is attributed to its low-pass filtering effects. However, current researches lack a study of how different signal components contributes to recommendations, and how to design strategies to properly use them well. To this end, from the view of spectral transformation, we analyze the important factors that a graph filter should consider to achieve better performance. Based on the discoveries, we design JGCF, an efficient and effective method for CF based on Jacobi polynomial bases and frequency decomposition strategies. Extensive experiments on four widely used public datasets show the effectiveness and efficiency of the proposed methods, which brings at most 27.06\% performance gain on Alibaba-iFashion. Besides, the experimental results also show that JGCF is better at handling sparse datasets, which shows potential in making recommendations for cold-start users.
\end{abstract}

\maketitle

% \section{Introduction}

% \begin{figure}
%     \centering
%     \includegraphics{}
%     \caption{}
%     \label{fig:my_label}
% \end{figure}

\section{Introduction}

Recommender systems are essential for online marketing platforms in providing users with relevant information and promoting products for content providers. Collaborative filtering~(CF), one of the cornerstones in the recommender system, has been prevailing in both the academic and industry for tens of years. The key motivation of CF is to mine the relationship between similar users and items for recommendations. One of the most successful CF methods is matrix factorization~(MF)~\cite{Koren2009MatrixFT}, which uses the inner product as the rating function and factorizes the user-item rating matrix into user embeddings and item embeddings. Despite its success in providing sufficient recommendations, only a linear operator is considered in processing user behaviors and the model capacity is less expressive. To overcome this difficulty, many recent studies have been proposed to introduce different non-linearity operations into CF, including multi-layer neural networks~\cite{Covington2016DeepNN,he2017neural}, recurrent networks~\cite{Hidasi2015SessionbasedRW,Wu2017RecurrentRN}, memory networks~\cite{Ebesu2018CollaborativeMN} and attention-based models~\cite{Chen2017AttentiveCF,Sun2019BERT4RecSR}. However, these models still ignore the high-order information on the user-item bipartite graph which may lead to information loss and degrade the final performance.

Recently, graph-based models~\cite{wang2019neural,he2020lightgcn,wu2021self} have arisen and become a popular choice for collaborative filtering. Their promising achievements rely on the effectiveness of mining the high-order neighborhood information in the user-item bipartite graphs. Current research on graph-based models for collaborative filtering can be categorized into two categories. The first kind of study learns the user embeddings and item embeddings by designing graph filters and conducting propagation to utilize high-order information and trains the user and item embeddings through mini-batch optimization. The other type of work directly derives the embeddings induced by the top-K singular values or eigenvalues~\cite{10.1145/3459637.3482264,10.1145/3477495.3532014,Peng2022SVDGCNAS} through SVD or eigendecomposition. On both lines of researches, current graph collaborative filtering uses either a low-pass filter or a band-stop filter to introduce the low-frequency or the high-frequency components, while lacking a discussion on how different components should be utilized. Besides, many propagation methods of current models depend on the monomial bases~(e.g. LightGCN), while there are some problems in monomial bases when approximating given functions. One is that they are not orthogonal with weight function and may have optimization issues~\cite{askey1985some}. The other is that they are low-pass filters that do not consider high-frequency signals and do not handle the middle-frequency signal well.

% This line of research does not need to train the embedding but it may take much time to conduct SVD or eigen-decomposition, and can not scale for large graphs. Besides, they may take longer inference time as they need dense matrix multiplication. 

To overcome the aforementioned challenges, we first analyze graph collaborative filtering from a spectral transformation view. We find that the low-frequency and high-frequency signals of the interactions in training data have strong linear correlations with those in testing data. While the middle-frequency signals do not have such strong linear correlations. This shows that a good graph filter should be able to model the two signal parts separately, one by a band-stop filter which uses simple linear models to avoid over-fitting, and the other part uses non-linear models to enhance the model capacity. Motivated by this analysis result, we utilize a general form of orthogonal polynomial bases named Jacobi polynomials to approximate the graph convolution. We further propose a signal decomposition strategy to model the two signal parts separately. By separating the low-frequency and high-frequency components from the middle-frequency component, we can use linear transformations to handle the linear correlation part (i.e., the low-frequency and the high-frequency components) and use another non-linear transformation to handle the complex middle-frequency component. Extensive experiments are conducted on 4 public datasets. The results show that JGCF is both effective~(up to 27.06\% performance gain) and efficient in collaborative filtering, and is good at handling sparse datasets. In summary, our contributions are:
\begin{itemize}
    \item We study the recommendation task from a spectral transformation view and identify the insightful finding as follows. We find that both low-frequency signals and high-frequency signals of interactions in training data have strong linear correlations with that in testing data, which are easy to fit by linear mapping. On the contrary, the middle-frequency signals do not have strong linear correlations.
    \item Based on this, we propose JGCF, a graph collaborative filtering method based on a filter designed under frequency decomposition strategy, and use a kind of orthogonal polynomial basis, Jacobi polynomials to approximate the filter.
    \item We conduct extensive experiments with state-of-the-art methods on 4 public benchmarks. The results verify the effectiveness and efficiency of JGCF. Besides, the results show that JGCF has the potential in handling cold-start users and items.
\end{itemize}

\section{Preliminary}

\subsection{Collaborative Filtering}

We denote the user set as $\mathcal{U}$, the item set as $\mathcal{I}$, and the number of all users and items by $N=|\mathcal{U}|+|\mathcal{I}|$. In collaborative filtering, the rating matrix is $\mathbf{R}\in\mathbb{R}^{|\mathcal{U}|\times |\mathcal{I}|}$ where $\mathbf{R}_{ij}=1$ if the $i$-th user and the $j$-th item have an interaction. Considering user $u$, item $i$ and the rating matrix $\mathbf{R}$, the collaborative filtering method aims to learn a score function $s(u,i,\mathbf{R})$ to rank candidate items the user may be interested in the future. For neural-based collaborative filtering, usually, a learnable embedding table of users and items will be used and is denoted by $\mathbf{E}^{(0)}\in\mathbb{R}^{N\times d}$. 

% Here $d$ is the dimension of embedding vectors.

To utilize graph-based methods for collaborative filtering, an adjacency matrix of the user-item bipartite graph based on $\mathbf{R}$ is usually required:
\begin{equation}
    \mathbf{A}=\begin{bmatrix} \mathbf{0} & \mathbf{R} \\ \mathbf{R}^T & \mathbf{0} \end{bmatrix}
\end{equation}
\noindent where $\mathbf{A}\in\mathbb{R}^{N\times N}$. To ensure training stability, most methods adopt the normalized adjacency matrix $\hat{\mathbf{A}}=\mathbf{D}^{-\frac{1}{2}}\mathbf{A}\mathbf{D}^{-\frac{1}{2}}$, where $\mathbf{D}$ is the diagonal matrix with $\mathbf{D}_{ii}=\sum_j\mathbf{A}_{ij}$ denoting the degree of node $i$. Each user and each item is further assigned with a learnable embedding $\mathbf{e}_u^{(0)}$ and $\mathbf{e}_v^{(0)}$ corresponding to their unique ID. Then graph neural network-based methods will incorporate high-order information through the propagation over neighbor nodes and usually, the representations of each layer are concatenated as the final node representation. For example, the propagation of the representative method, LightGCN, can be formulated as:
\begin{equation}
    \mathbf{E}^{(k+1)}=\hat{\mathbf{A}}\mathbf{E}^{(k)}
\end{equation}
\noindent where $k$ denotes the $k$-th layer GNN, and the final node embedding is obtained by the weighted average of embeddings in different layers:
\begin{equation}
    \mathbf{E}=\sum_{k=0}^{K}\mathbf{\alpha}_k\mathbf{E}^{(k)}.
    \label{eq:average}
\end{equation}
The smoothed embedding table $\mathbf{E}$ can be further utilized to compute the recommendation scores for different user-item interactions.
\subsection{Spectral Graph Neural Network}

Spectral GNN conducts graph convolutions in the domain of the Laplacian spectrum. The graph Laplacian matrix can be derived by $\mathbf{L}=\mathbf{D}-\mathbf{A}$. Its normalized version is $\hat{\mathbf{L}}=\mathbf{I}-\hat{\mathbf{A}}$. As $\mathbf{L}$ is a symmetric positive definite matrix, we can obtain its eigenvalue and corresponding eigenvector by eigendecomposition: 
\begin{equation}
    \mathbf{L}=\tilde{\mathbf{U}}\tilde{\mathbf{\Lambda}}\tilde{\mathbf{U}}^T
\end{equation}  
\noindent where $\tilde{\mathbf{U}}\in\mathbb{R}^{N\times N}$ is the eigenvector and $\tilde{\mathbf{\Lambda}}\in\mathbb{R}^{N\times N}$ is a diagonal matrix with eigenvalues on the diagonal. The graph Fourier transform of a signal $x\in\mathbb{R}^{N}$ is then defined as $\hat{x}=\tilde{\mathbf{U}}^Tx$ and the inverse transformation is $x=\tilde{\mathbf{U}}\hat{x}$. The transform enables the formulation of operations such as filtering in the spectral domain. The filtering operation on signals $x$ with a filter $g_\theta$ is defined as:
\begin{equation}
y=g_\theta(\mathbf{L})x=g_\theta(\tilde{\mathbf{U}}\tilde{\mathbf{\Lambda}}\tilde{\mathbf{U}}^T)x=\tilde{\mathbf{U}}g_\theta(\tilde{\mathbf{\Lambda}})\tilde{\mathbf{U}}^Tx.
\end{equation}
Directly computing $g_\theta(\tilde{\mathbf{\Lambda}})$ is infeasible as the decomposition of the Laplacian matrix is time-consuming, especially for large-scale graphs. Therefore, spectral GNN usually uses some polynomial bases to approximate $g(\tilde{\mathbf{\Lambda}})$ like:
\begin{equation}
    g_\theta(\tilde{\mathbf{\Lambda}})=\sum_{k=0}^{K}\theta_k\mathbf{P}_k(\mathbf{\tilde{\Lambda}})
\end{equation}
\noindent where $\theta_k$ is a learnable or fixed scalar for the $k$-th order polynomial. Monomials, Chebyshev polynomials, and Bernstein polynomials are some typical choices in the approximation. The filtered signals can be deployed for downstream tasks.

\section{Methodology}

\subsection{Design Motivation.}

\subsubsection{\textbf{A Spectral Transformation View of Recommendation}}
Given existing user-item interactions, the target of recommender systems is to predict the missing ones, and this task can be viewed as learning a mapping function that transforms previous interactions into possible links between users and items in the future. Mathematically, there is an adjacency matrix $\mathbf{A}\in\mathbb{R}^{N\times N}$ that describes historical user-item interactions, with $\mathbf{A}_{ij}=1$ when user $i$ already has clicked ~(or other pre-defined behaviors) item $j$. $\mathbf{B}\in\mathbb{R}^{N\times N}$ is the adjacency matrix of missing user-item interactions where $\mathbf{B}_{ij}=1$ indicates that user $i$ will have an interaction with item $j$. Usually, in industrial scenarios, interactions of $\mathbf{B}$ are regarded to occur later than those of $\mathbf{A}$. Then the recommendation model can be considered as a mapping $f(\cdot)$ from the existing interactions to the future possible ones. Given $f(\cdot)$ a matrix polynomial, the optimization of the transformation $f(\cdot)$ can be formulated as~\cite{Kunegis2009LearningSG}:
\begin{equation}
    \mathop{\text{minimize}}_{f\in\mathcal{H}}||f(\mathbf{A})-\mathbf{B}||_F
    \label{eq:tranformation}
\end{equation}
\noindent where $\mathcal{H}$ is the hypothesis space of all possible transformations and $||\cdot||_F$ is the Frobenius norm. As $\mathbf{A}$ is a real symmetric matrix, through the eigendecomposition, we have $\mathbf{A}=\mathbf{U}\mathbf{\Lambda}\mathbf{U}^T$  where $\mathbf{U}\in\mathbb{R}^{N\times N}$ is a group of orthogonal eigenvectors and $\mathbf{\Lambda}\in\mathbb{R}^{N\times N}$ is a diagonal matrix with all corresponding eigenvalues on the diagonal. Given the assumption that $f$ is a matrix polynomial, Eq~(\ref{eq:tranformation})  can be rewrite as:
\begin{equation}
\begin{split}
    ||f(\mathbf{A})-\mathbf{B}||_F&=||f(\mathbf{U}\mathbf{\Lambda}\mathbf{U}^T)-\mathbf{B}||_F\\
    &=||\mathbf{U}f(\mathbf{\Lambda})\mathbf{U}^T-\mathbf{B}||_F \\
    &=||f(\mathbf{\Lambda})-\mathbf{U}^T\mathbf{B}\mathbf{U}||_F.
    \label{eq:rewrite}
\end{split}
\end{equation}
\noindent Eq~(\ref{eq:rewrite}) is derived based on the fact that the Frobenius norm is invariant with the multiplication of an orthogonal matrix on the right and its transpose on the left. \autoref{eq:rewrite}  can be decomposed into two parts: the Frobenius norm of off-diagonal entries in $f(\mathbf{\Lambda})-\mathbf{U}\mathbf{B}\mathbf{U}^T$ that is independent of $f$, and the Frobenius norm of its diagonal entries. Then it leads to the following equivalent objective:
% least-squares problem equivalent to Eq~(\ref{eq:tranformation}).
\begin{equation}
    \mathop{\text{minimize}}_{f\in\mathcal{H}}\sum_i\left ( f(\mathbf{\Lambda}_{ii})-\mathbf{U}^T_{\cdot i}\mathbf{B}\mathbf{U}_{\cdot i} \right )^2,
\end{equation}
\noindent which is a one-dimensional least-squares curve fitting problem of size $N$. In the following section, we will try to find out the properties of $f(\cdot)$ by observing the relationship between the eigenvalues and the diagonal of the transformed test interactions.  

\subsubsection{\textbf{Observations of Real World Dataset}}
\begin{figure}
    \centering
    \subfigure{\includegraphics[width=.48\linewidth]{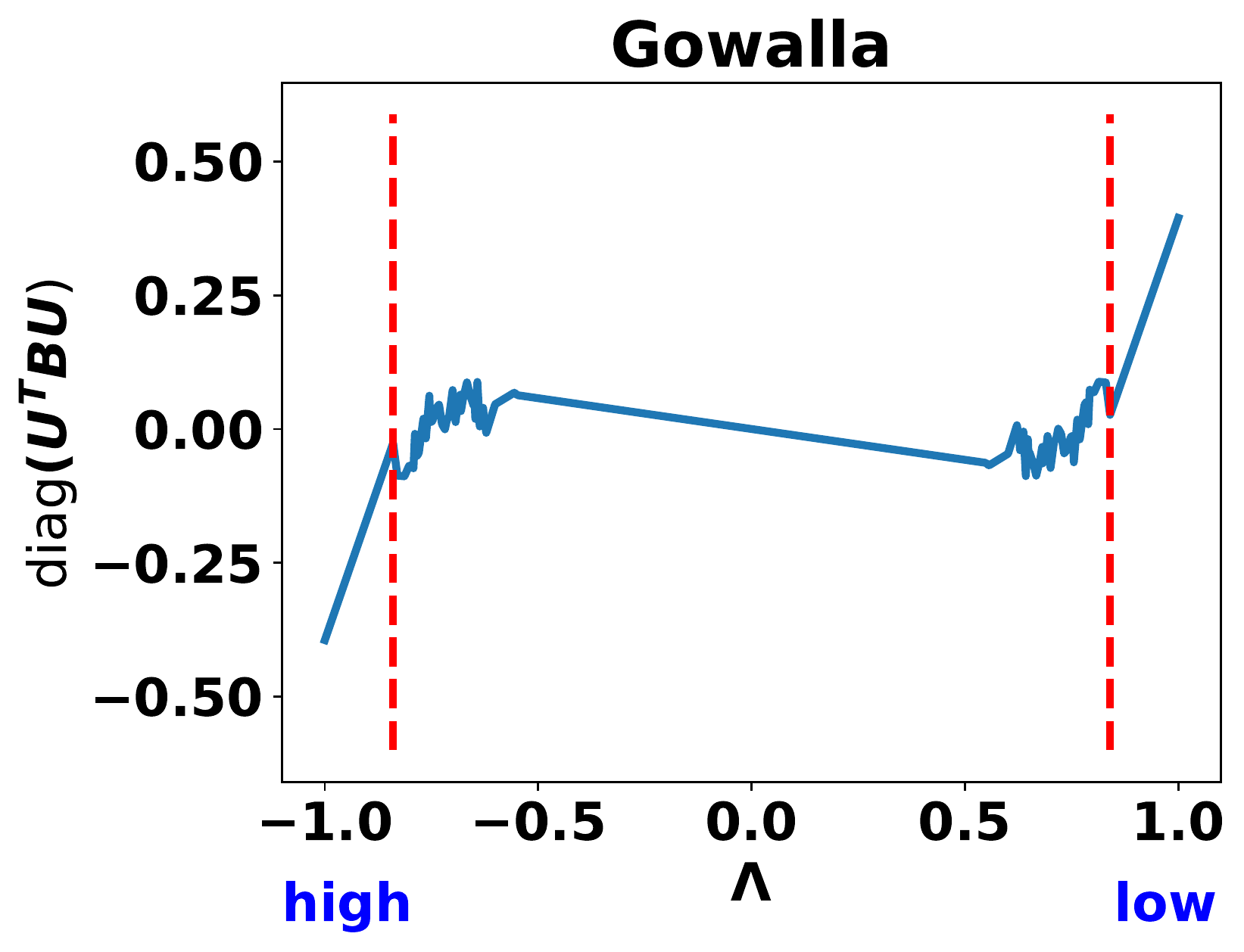}}
    \subfigure{\includegraphics[width=.48\linewidth]{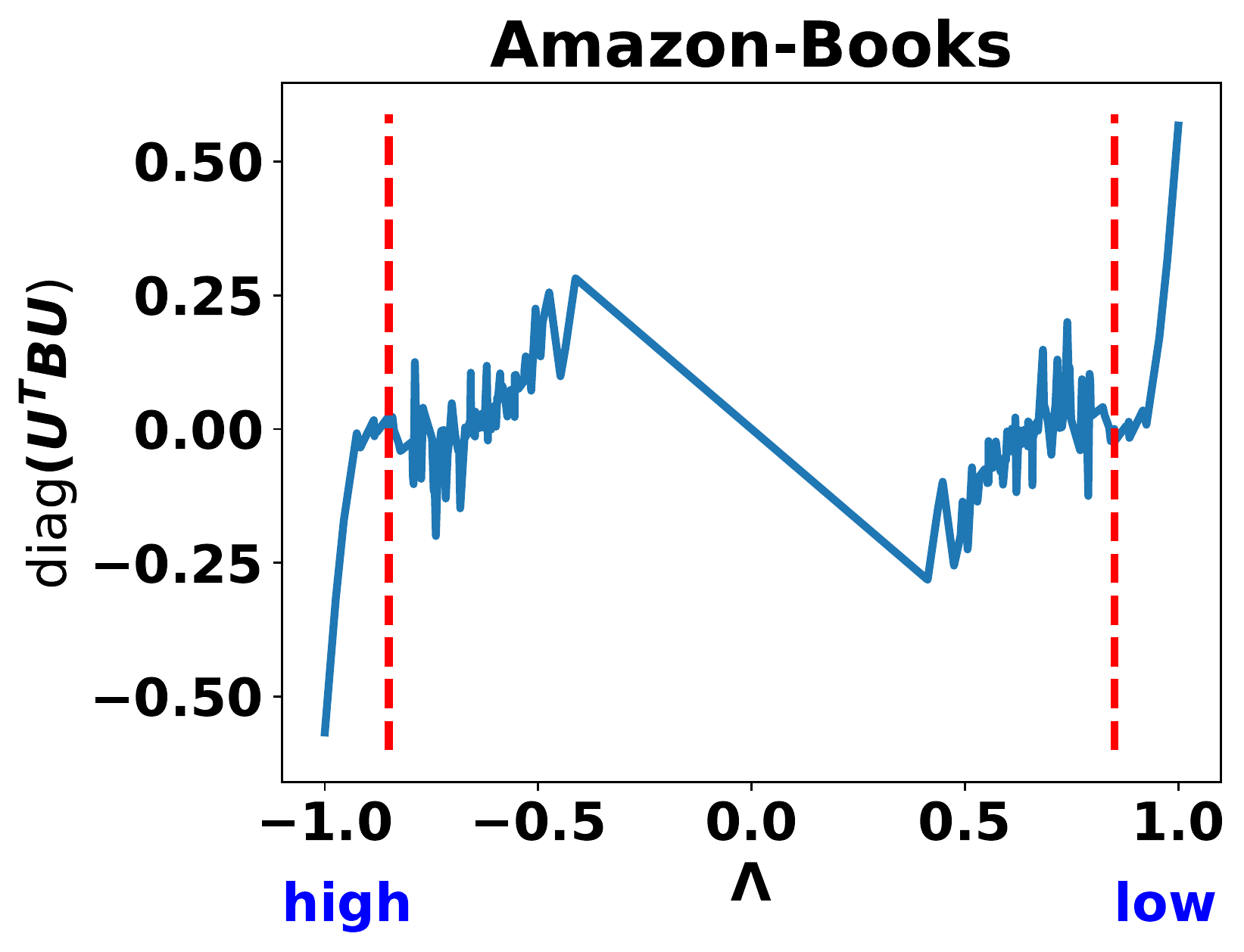}}
    \subfigure{\includegraphics[width=.48\linewidth]{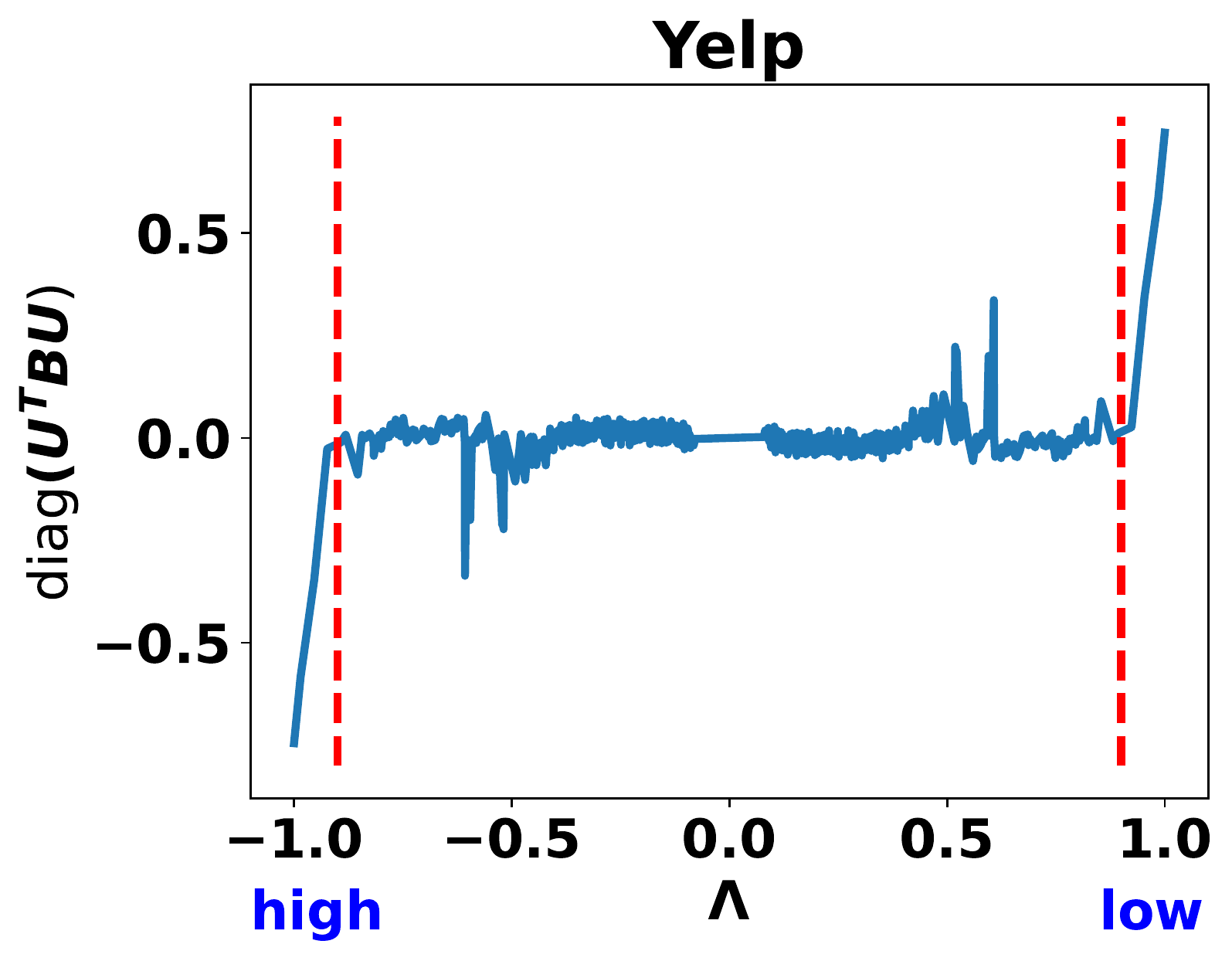}}
    \subfigure{\includegraphics[width=.48\linewidth]{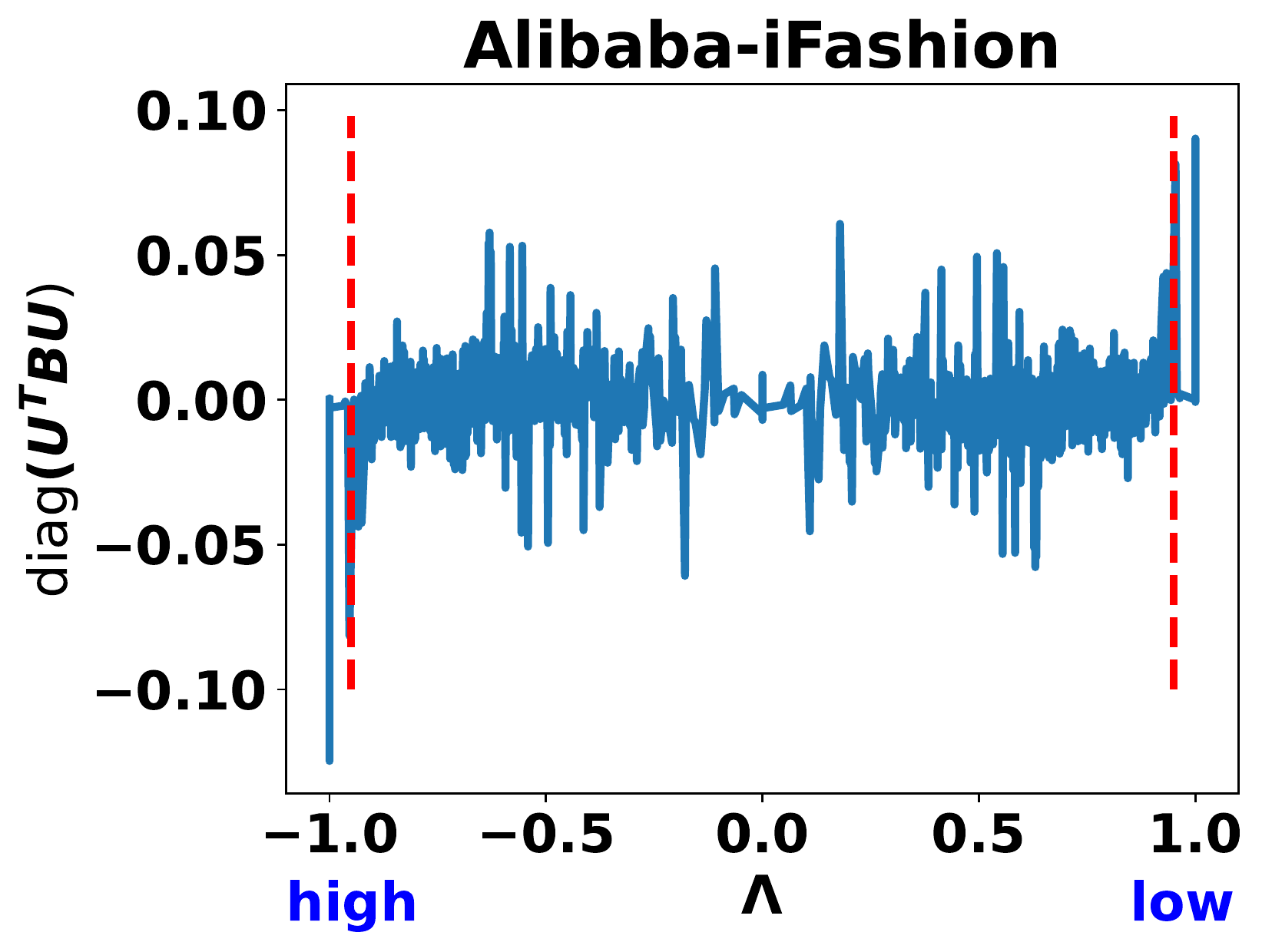}}
    \caption{Relation between the eigenvalues $\mathbf{\Lambda}$ of $\hat{\textbf{A}}$ and the diagonal of $\mathbf{U}^T\mathbf{B}\mathbf{U}$. Larger eigenvalue $\mathbf{\Lambda}_{ii}$ indicates lower frequency signals.}
    % \lun{Need more explanation for $\mathbf{\Lambda}$ and $\mathbf{U}^T\mathbf{B}\mathbf{U}$.} }
    \label{fig:relation}
    \vspace{-0.2cm}
\end{figure}
The relationship between the diagonals of $\mathbf{\Lambda}$ and $\mathbf{U}^T\mathbf{B}\mathbf{U}$ on four real-world datasets are shown in Figure~\ref{fig:relation}. Due to the large scale of datasets that causes a long time for eigendecomposition, BFS starting from some randomly selected nodes is conducted to reduce the size of user-item graphs.
% we sample on the user-item graph to reduce the scale by BFS starting of several randomly selected initial nodes.
From the figure we have the following discoveries:
\begin{enumerate}
    \item The low-frequency component~(values near 1.0) and high-frequency component~(values near -1.0) are positively correlated to the diagonal of $\mathbf{U}^T\mathbf{B}\mathbf{U}$, indicating that we can preserve information of these two components by relatively simple models so that patterns from historical interactions can be easily promoted to the unseen ones.
    \item The mid-frequency component~(values around 0.0) of the training graph is less correlated with signals on the test graph, meaning that signals within this band contain less information about the user interest. 
\end{enumerate}
\subsubsection{\textbf{Inspiration for the Filter Design.}}  
From the above observations, we can conclude some key properties a graph filter should have for better recommendations. The filter should behave like a band-stop filter that preserves the low-frequency component and the high-frequency component, as they are linearly correlated with the future possible interactions and are easy to fit. At the same time, as the correlations of middle-frequency are quite intricate, the filter should either suppress such signals or use some non-linear functions to transform the middle-frequency signal to the target spectrum, i.e., $\text{diag}(\mathbf{U}^T\mathbf{B}\mathbf{U})$. Besides in Figure~\ref{fig:relation}, we find the filer should have positive responses at the low-frequency area, while having negative responses at the high-frequency area. 

\subsubsection{\textbf{Polynomial Approximation View of LightGCN}} 

From the above observation and inspirations, we revisit the polynomial form of LightGCN, which can be organized from the perspective of spectral GNN as:
\begin{equation}
    \mathbf{E}=\frac{1}{K+1}\sum_{k=0}^{K}\mathbf{E}^{(k)}=\sum_{k=0}^{K}\frac{\hat{\mathbf{A}}^k}{K+1}\mathbf{E}^{(0)},
\end{equation}
\noindent where a series of monomial bases is applied to approximate the filter. However, the monomials basis may raise several issues. One is that they are not orthogonal to each other with weight functions. The other is that its ability in suppressing middle-frequency signals is limited. To verify this conclusion, We illustrate the signal response of the Monomial basis in Figure~\ref{fig:filter}. The response values are normalized by $\frac{\sum_{k=0}^K\alpha_k\lambda_t^k}{\sum_{k=0}^K\alpha_k}$~(i.e., a ratio to the maximum). The figure reveals that although it is a low-pass filter that emphasizes the low-frequency signals~(eigenvalue around 1.0), it does not significantly suppress the middle-frequency signals. Moreover, it does not produce negative responses for the high-frequency component~(eigenvalue around -1.0), which can not meet the requirement of a desirable graph filter and will limit its expressive power.

\subsection{Jacobi Polynomial Bases}

\subsubsection{\textbf{Introduction}} To overcome the limitations of LightGCN, we should consider another solution to approximate the graph signal filter.
% a group of more appropriate polynomial bases as the graph signal filter. 
One naive way is to decompose the adjacency matrix and directly operate on eigenvalues. However, the complexity is $\mathcal{O}(N^2)$ and it is hard to scale to large graphs. Another choice is to find new polynomial bases with some requirements. Typically, the polynomial bases should be orthogonal to guarantee convergence. Moreover, the polynomial should be expressive enough so that it can approximate the filters of different user-item graphs. Jacobi polynomial is one suitable candidate that is widely used in numerical analysis and signal processing~\cite{askey1985some}. Chebyshev and Legendre polynomials can be seen as its special cases. The Jacobi basis $\mathbf{P}^{a,b}_k(x)$ has the following form that when $k=0$ or $k=1$:
\begin{equation}
    \begin{split}
        \mathbf{P}_0^{a,b}(x)&=1, \\
        \mathbf{P}_1^{a,b}(x)&=\frac{a-b}{2}+\frac{a+b+2}{2}x.
    \end{split}
\end{equation}
\noindent For $k\ge 2$, it satisfies
\begin{equation}
    \mathbf{P}_k^{a,b}(x)=(\theta_k z+\theta_k')\mathbf{P}_{k-1}^{a,b}(x)-\theta_k^{''}\mathbf{P}_{k-2}^{a,b}(x),
\end{equation}
\noindent where
\begin{equation}
\begin{split}
    \theta_k &=\frac{(2k+a+b)(2k+a+b-1)}{2k(k+a+b)},\\
    \theta_k^{'} &=\frac{(2k+a+b-1)(a^2-b^2)}{2k(k+a+b)(2k+a+b-2)},\\
    \theta_k^{''} &=\frac{(k+a-1)(k+b-1)(2k+a+b)}{k(k+a+b)(2k+a+b-2)}.\\ 
\end{split}
\label{eq:theta}
\end{equation}
\noindent In which $\mathbf{P}_k^{a,b}(x), k=0,...,K$ are orthogonal with weight function $(1-\lambda)^a(1+\lambda)^b$ on $[-1,1]$ with $a,b>{-1}$. Jacobi polynomials provide a general solution for graph signal filtering. In more detail, we can adjust $a$ and $b$ to control the signal filter for various graphs. To further demonstrate the influence of $a$ and $b$, we plot the signal responses of Jacobi polynomials under different $a$ and $b$ combinations in Figure~\ref{fig:a_and_b}. We find that the relative scale of $a$ and $b$ affect the signal response of the filter. When $a$ is fixed, larger $b$ indicates that the filter will enhance the response of the high-frequency signals. Thus, we can tune $a$ and $b$ to meet the need of different graphs. 
% Besides, the Jacobi polynomial is a type of second-order polynomial that ensures a faster convergence when approximating a given function~\cite{berthier2020accelerated}. 
With the above properties, the \textbf{J}acobi polynomial-based \textbf{G}raph \textbf{C}ollaborative \textbf{F}iltering (JGCF) is supposed to be effective for the user-item graph.

\subsubsection{\textbf{Signal Response of Jacobi Polynomials}} 
\begin{figure}
    \centering
    \subfigure{\includegraphics[width=.38\linewidth]{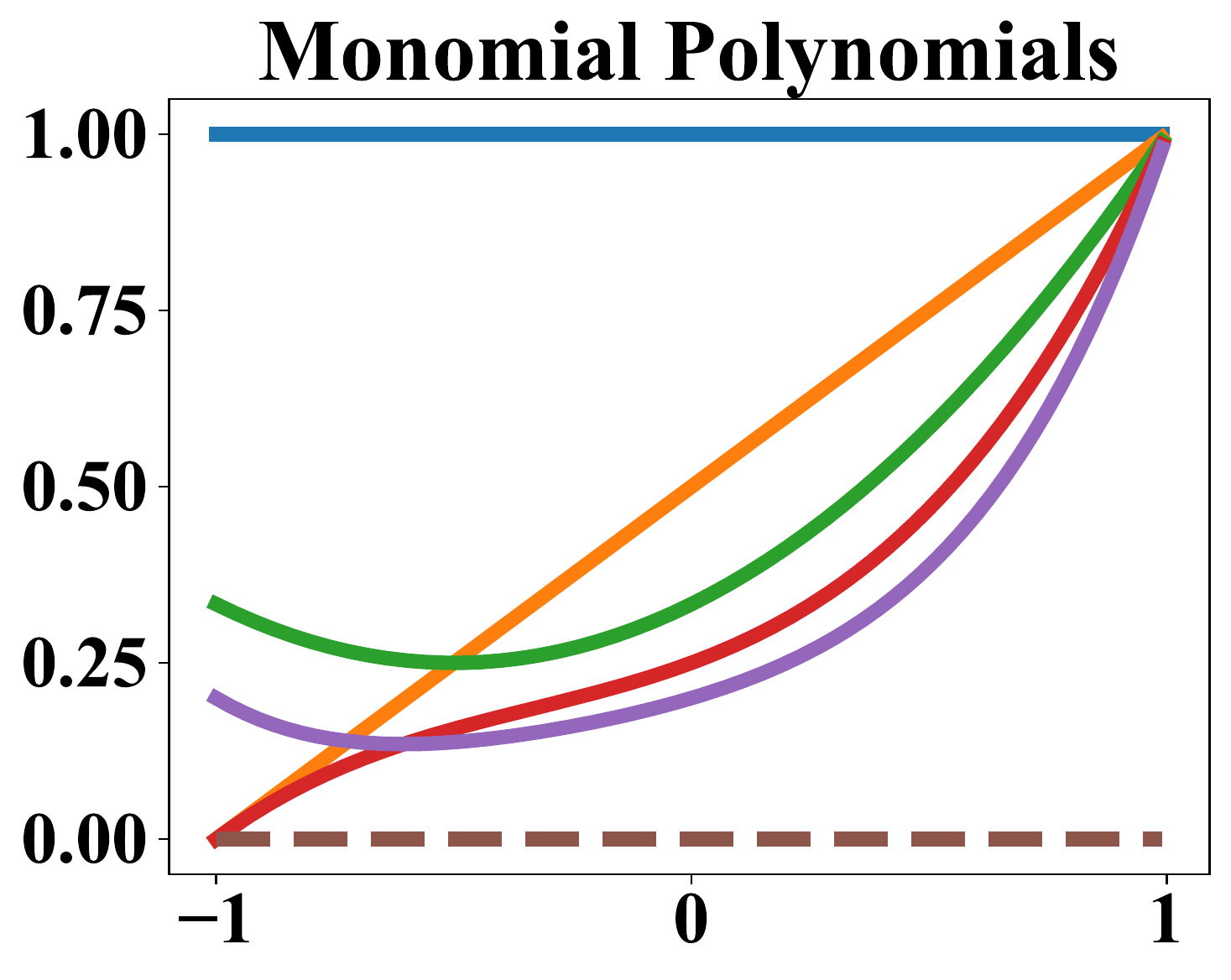}}
    \subfigure{\includegraphics[width=.54\linewidth]{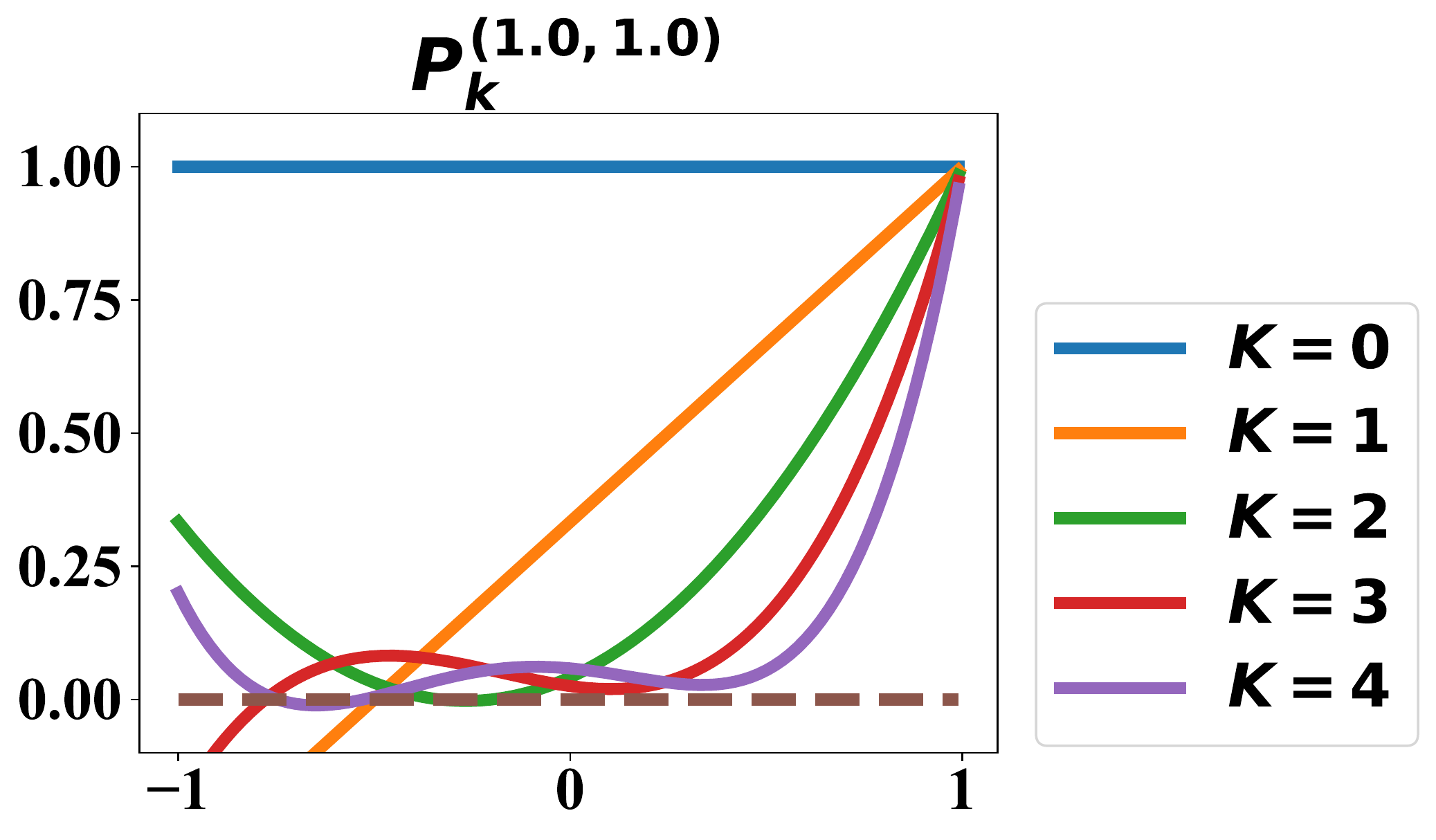}}
    \subfigure{\includegraphics[width=.54\linewidth]{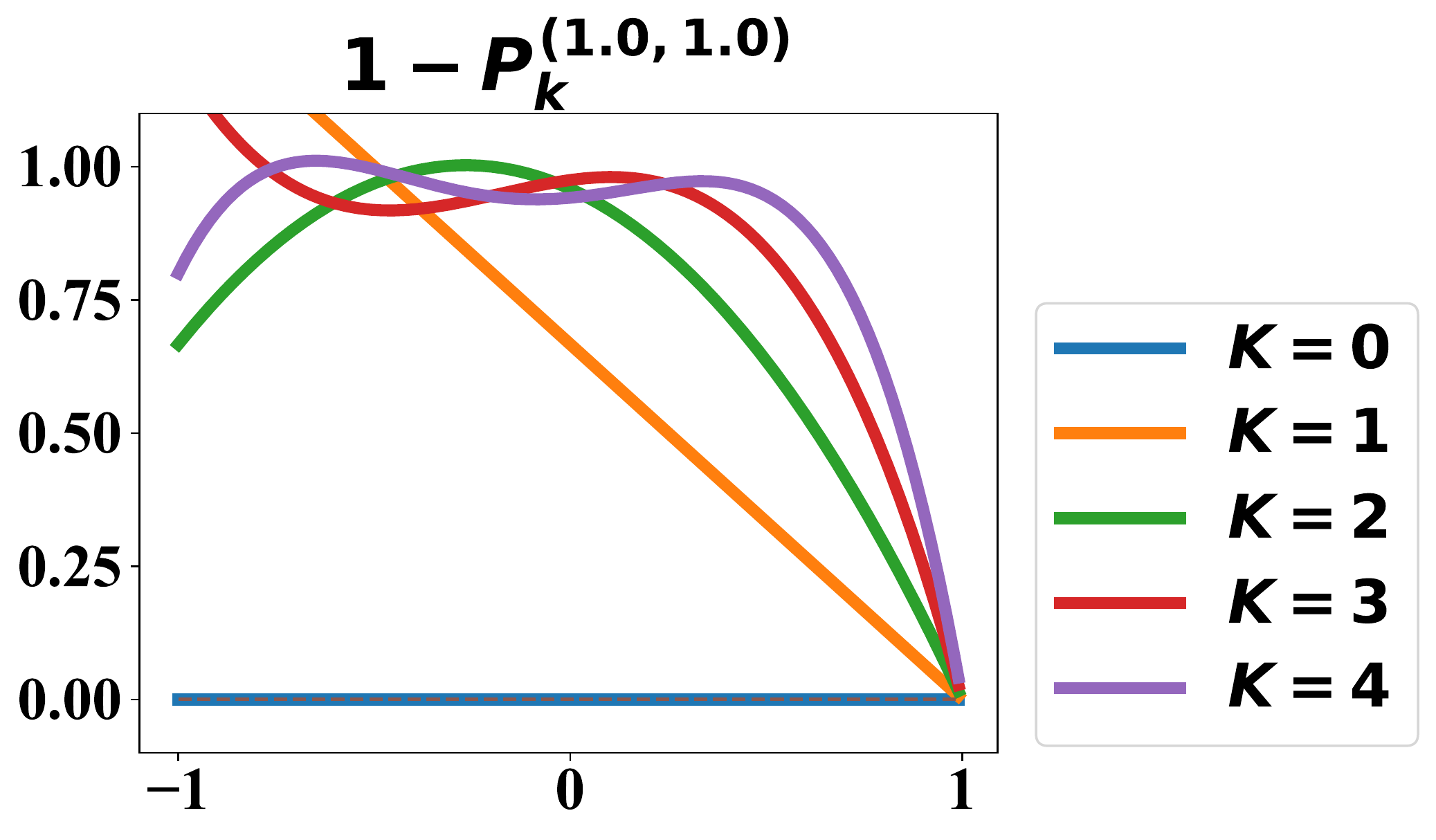}}
    \caption{Signal response for filters of LightGCN~(upper left), JGCF-band stop~(upper right) and JGCF-band pass~(bottom) with $a=1.0$ and $b=1.0$ under different orders.}
    \label{fig:filter}
\vspace{-0.2cm}
\end{figure}

The signal response of Jacobi polynomial is also displayed in Figure~\ref{fig:filter}. Following the above discussion we can conclude that the signal response of JGCF is more well-suited compared with the one of LightGCN. Firstly it suppresses the middle-frequency signal and enhances both the low-frequency signal and high-frequency signal. Besides, when $K$ is odd, it produces negative responses for the high-frequency component, which is aligned with the motivation from Figure~\ref{fig:relation}. 

\subsubsection{\textbf{Correlation Comparison}} To further verify the effectiveness of the Jacobi polynomial, we replace the function $f(\cdot)$ in Eq~(\ref{eq:rewrite}) with commonly used polynomials and show the Pearson correlation between the transformed spectrum and the diagonals of $\mathbf{U}^T\mathbf{B}\mathbf{U}$. We keep $K=3$ following common GNN settings. The result is shown in Table~\ref{tab:pearson}. We can discover that the Jacobi polynomial with $a=b=1.0$ achieves the strongest Pearson correlation with the diagonals of $\mathbf{U}^T\mathbf{B}\mathbf{U}$, and distinctly outperforms other polynomials. Such results reveal great potential of the Jacobi polynomial in handling the target dataset by adjusting $a$ and $b$. 
\begin{table}[t]
\Large
    \centering
    \caption{Pearson correlation coefficient between transformed spectrum $f(\mathbf{\Lambda})$ and the diagonal of $\mathbf{U}^T\mathbf{B}\mathbf{U}$. A larger value means stronger positive linear correlations.}
    \resizebox{.9\linewidth}{!}{
    \begin{tabular}{c|cccc}
    \toprule
      & Monomial & Chebyshev & BernNet & Jacobi \\
    \midrule
       Gowalla  & 0.0746 & 0.2602 &  0.2217 & \textbf{0.6966} \\
       Amazon-Books & 0.0788 & 0.2665 & 0.2293 & \textbf{0.5935} \\
       Yelp & 0.0807 & 0.3011 & 0.2639 & \textbf{0.4358} \\
       Alibaba-iFashion & 0.0310 & 0.1037 & 0.0901 & \textbf{0.2123} \\
    \bottomrule
    \end{tabular}}
    \label{tab:pearson}
\end{table}

\subsection{Graph Signal Filtering via Frequency Decomposition}
Despite choosing the proper polynomial bases, we still need different designs for components in different frequencies. As we mention, the filter based on Jacobi polynomial bases behaves like a band-stop filter that emphasizes both the low-frequency and the high-frequency signals while suppressing the middle-frequency signal. The low-frequency and the high-frequency signal are easy to fit, so we do not use any activation function and the node representation is obtained by averaging embeddings of different orders:
\begin{equation}
\begin{split}
    \mathbf{E}_{\text{band-stop}}^{(K)}&=g_K(\mathbf{I}-\mathbf{L})\mathbf{E}^{(0)}=\mathbf{U}g_K(\mathbf{\Lambda})\mathbf{U}^T\mathbf{E}^{(0)} \\
    &=\frac{1}{K+1}\sum_{k=0}^{K}\mathbf{P}_k^{a,b}(\hat{\mathbf{A}})\mathbf{E}^{(0)},
\end{split}
\end{equation}
% \begin{equation}
% \begin{split}
%     \mathbf{E}_{\text{band-stop}}^{(K)}&=\mathbf{U}g_K(\mathbf{\Lambda})\mathbf{U}^T\mathbf{E}^{(0)}=\mathbf{} \\
%     &=\frac{1}{K+1}\sum_{k=0}^{K}\mathbf{P}_k^{a,b}(\hat{\mathbf{A}})\mathbf{E}^{(0)} 
% \end{split}
% \end{equation}
\noindent and for the mid-frequency component, a typical design is $\alpha\mathbf{I}-\mathbf{U}g_K(\mathbf{\Lambda})\mathbf{U}^T$. $\alpha$ is a coefficient that controls the impact of the middle-frequency signal. The signal response of this filter is illustrated in Figure~\ref{fig:filter}. It shows that the filter is similar to a band-pass filter that emphasizes the middle-frequency components. Hence, the representation from the band-pass filter is formulated as:
\begin{equation}
\begin{split}
\mathbf{E}_{\text{band-pass}}^{(K)}&=\text{tanh}\left(\left(\alpha\textbf{I}-\mathbf{U}g_K(\mathbf{\Lambda})\mathbf{U}^T\right) \mathbf{E}^{(k)}\right ) \\
    &=\text{tanh}\left (\left(\alpha\mathbf{I}-\frac{1}{K+1}\sum_{k=0}^{K}\mathbf{P}_k^{a,b}(\hat{\mathbf{A}})\right )\mathbf{E}^{(0)}\right )
\end{split}
\end{equation}
\noindent The final node representation is the concatenation of them: 
\begin{equation}
    \mathbf{E}=[\mathbf{E}_{\text{band-stop}}^{(K)};\mathbf{E}_{\text{band-pass}}^{(K)}]
\end{equation}
\begin{figure}[t]
    \centering
    \includegraphics[width=.9\linewidth]{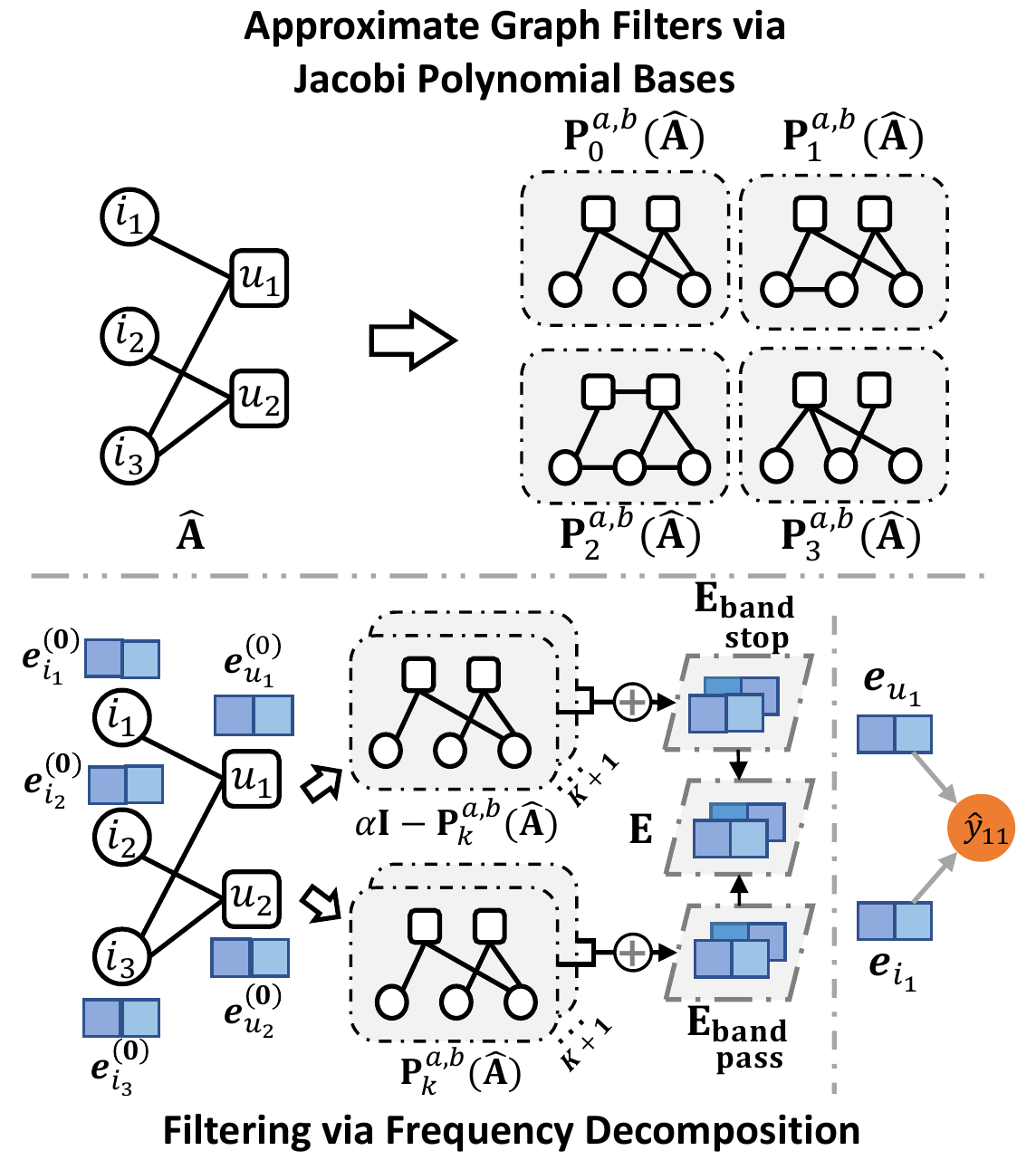}
    \caption{Overview of the Jacobi Graph Collaborative Filtering. It uses a group of Jacobi polynomial bases to construct a bank of filters. Then it applies a frequency decomposition strategy to process different signal components. The filtered embeddings are then used for prediction.}
    \label{fig:framework}
\end{figure}

\subsection{Jacobi Graph Collaborative Filtering}

In this section, we elaborate on the proposed Jacobi graph collaborative filtering~(JGCF) that includes 4 major modules: 1) embedding initialization; 2) graph filtering; 3) prediction and 4) optimization. The overall framework is shown in Figure~\ref{fig:framework}. 
\subsubsection{\textbf{Embedding Initialization}} 
When user IDs and item IDs are available without other features, each user and item is assigned with a randomly initialized learnable embedding:
\begin{equation}
    \mathbf{E}^{(0)}=[\mathbf{e}_1^{(0)},...,\mathbf{e}_{|\mathcal{U}|}^{(0)},\mathbf{e}_{|\mathcal{U}|+1}^{(0)},...\mathbf{e}_{|\mathcal{U}|+|\mathcal{I}|}^{(0)}].
\end{equation}
\noindent For applications where features are available, we can transform the features and concatenate them with the embeddings.
\subsubsection{\textbf{Graph Filtering}} 
We use a recursive design to filter the graph signals. For $k=1$ we have:
\begin{equation}
\mathbf{E}^{(1)}=\mathbf{P}_{1}^{a,b}(\hat{\mathbf{A}})\mathbf{E}^{(0)}=\frac{a-b}{2}+\frac{a+b+2}{2}\hat{\textbf{A}}\mathbf{E}^{(0)},
\end{equation}
\noindent and when $k\ge 2$ it should be like:
\begin{equation}
\begin{split}
\mathbf{E}^{(k)}=\mathbf{P}^{a,b}_k(\hat{\textbf{A}})\mathbf{E}^{(0)}=&\theta_k\hat{\textbf{A}}\mathbf{P}_{k-1}^{a,b}(\hat{\mathbf{A}})\mathbf{E}^{(0)}+\theta_k'\mathbf{P}_{k-1}^{a,b}\mathbf{E}^{(0)}\\ 
    &-\theta''_k\mathbf{P}_{k-2}^{a,b}(\hat{\textbf{A}})\mathbf{E}^{(0)}
\end{split}
\end{equation}
\noindent where $\theta_k$, $\theta_k'$ and $\theta_k''$ are derived from Eq.(\ref{eq:theta}). The final embedding is derived by the frequency decomposition strategy in Eq.(14)-(16). 
\subsubsection{\textbf{Prediction}}
Then we use the embedding after filtering to make recommendations. The user $u$'s interest score to an item $i$ is given by the inner product of the corresponding embeddings:
\begin{equation}
\hat{y}_{ui}=\mathbf{e}_u^T\mathbf{e}_i
\end{equation}
\noindent where $\hat{y}$ is the predicted score for user $u$ and item $i$. $\sigma(\cdot)$ is the Sigmoid function. $\hat{y}$ is then used as the ranking scores.
 
\subsubsection{\textbf{Optimization}} We then use the bayesian personalized ranking~(BPR)~\cite{rendle2009bpr}, a widely used matrix factorization~(MF) method for model training. BPR trains the user and item embeddings by encouraging the prediction score of the observed interactions to be higher than those that are not observed. Similar to~\cite{he2020lightgcn}, the only parameters that need to be updated are the embeddings of users and items $\mathbf{E}^{(0)}$, thus it enjoys efficient model training as MF. BPR has the following form:
\begin{equation}
    \mathcal{L}_{\text{BPR}}=-\sum_{u\in\mathcal{U}}\sum_{i\in\mathcal{N}_u}\sum_{j\notin\mathcal{N}_{u}}\text{ln}\sigma(\hat{y}_{ui}-\hat{y}_{uj})+\lambda ||\mathbf{E}^{(0)}||^2
\end{equation}
\noindent where $\mathcal{N}_{u}$ is the set of neighbors of user $u$ that user $u$ previously interacted. $\lambda$ is a coefficient that controls the influence of the $L_2$-regularization term. 

\subsection{Model Analysis and Discussion}

\subsubsection{\textbf{Time Complexity}}
The proposed JGCF is highly efficient. The main computation cost is the polynomial terms which can be pre-computed and directly used during training. Given $\mathcal{E}$ the edge set, $d$ the embedding dimension, the computation time for the $K$'th Jacobi Polynomial is $\mathcal{O}(K|\mathcal{E}|)$ and the total computational time for the graph convolution is $\mathcal{O}\left ((K|\mathcal{E}|+N)d\right )$, which is a linear combination to the total nodes and edges. For training, the complexity is the same as matrix factorization which uses a BPR loss. 

\subsubsection{\textbf{Polynomial Coefficients.}} Many SpectralGNN uses learnable polynomial coefficients as in Eq~(\ref{eq:average}). Our model does not use the learnable coefficients as we find uniformly assigning $\alpha_k$ to $\frac{1}{K+1}$ leads to better performance in general. Thus we do not use a special component to optimize it. Although carefully tuning it may lead to better performance, we do not do that so as to keep the model as simple and efficient as possible. For larger k, we multiply the representation by a discount factor to avoid over-smoothing.
\section{Experiments}

\subsection{Experimental Setup}

\subsubsection{\textbf{Datasets.}} We conduct experiments on widely used public datasets as follows.
\begin{itemize}
    \item \textbf{Gowalla.}~\cite{cho2011friendship} A point of interest network that is widely used in recommendation algorithm evaluation.
    \item \textbf{Amazon-Books.}~\cite{mcauley2015image} is a user rating history dataset based on the amazon book online store.
    \item \textbf{Yelp.}~\footnote{https://www.yelp.com/dataset/challenge} is a business recommendation dataset, where each business is viewed as an item.
    \item \textbf{Alibaba-iFashion.}~\cite{chen2019pog} includes clicks on the items and outfits from Taobao iFashion users.
\end{itemize}

The statistics of all the datasets are shown in Table~\ref{tab:dataset}. For each dataset, we randomly sample 80\% interactions of each user for training and 10\% of all the interactions for validation. The remaining 10\% interactions are used for testing. We sample one negative item for each positive instance to form the training set. 
\begin{table}[h]
\Large
    \centering
    \caption{Statistics of the Datasets.}
    \resizebox{.9\linewidth}{!}{
    \begin{tabular}{ccccc}
    \toprule
       \textbf{Datasets}  & \#\textbf{Users} & \#\textbf{Items} & \#\textbf{Interactions} & \textbf{Sparsity}  \\
    \midrule
       %  TenRec & 24,517 & 7,356 & 348,736 & 99.81\% \\
       Gowalla  & 29,858 &  40,981 & 1,027,464 & 99.92\% \\
       Amazon-Books & 52,642 &  91,598 & 2,984,108 & 99.94\% \\
       Yelp2018 &  31,668 & 38,048 & 1,561,406 & 99.87\% \\
       Alibaba-iFashion & 300,000 &  81,614 & 1,607,813 & 99.99\% \\
       \bottomrule
    \end{tabular}}
    \label{tab:dataset}
\end{table}

\subsubsection{\textbf{Baselines.}} We compare JGCF with recent CF baselines and also show the performance as the backbone of the self-supervised learning method. The baselines used are listed as follows.

Collaborative filtering baselines:

\begin{itemize}
    \item \textbf{BPR.}~\cite{rendle2009bpr} A matrix factorization framework based on bayesian empirical loss.
    \item \textbf{NeuMF.}~\cite{he2017neural} replaces the dot product in the MF model with a multi-layer perceptron to learn the match function of users and items.
    % \item \textbf{SpectralCF.}~\cite{10.1145/3240323.3240343} uses graph spectral filtering method to make recommendation.
    \item \textbf{NGCF.}~\cite{wang2019neural} adopts user-item bipartite graphs to incorporate high-order information and utilizes GNN to enhance CF.
    \item \textbf{DGCF.}~\cite{wang2020disentangled} produces disentangled representations for the user and item to improve the performance.
    \item \textbf{LightGCN.}~\cite{he2020lightgcn} simplifies the design of GCN to make it more concise and appropriate for recommendation.
    % \item \textbf{SimpleX.}~\cite{mao2021simplex} propose a simple but effective self-supervised learning method for collaborative filtering.
    \item \textbf{GTN.}~\cite{fan2022graph} propose to use graph trend filtering to denoise the user-item graph. 
    \item \textbf{RGCF.}~\cite{tian2022learning} learns to denoise the user-item graph by removing noisy edges and then adding edges to ensure diversity.
    \item \textbf{GDE.}~\cite{10.1145/3477495.3532014} uses embeddings of top-k eigenvalues of the user-user graph and item-item graph to make recommendation.
    \item \textbf{DirectAU.}~\cite{10.1145/3534678.3539253} A collaborative filtering method with representation alignment and uniformity.
\end{itemize}

Self-supervised learning baselines for backbone comparisons:

\begin{itemize}
    \item \textbf{SGL.}~\cite{wu2021self} introduces self-supervised learning to enhance recommendation. We adopt SGL-ED as the baseline.
    \item \textbf{NCL.}~\cite{lin2022improving} improves graph collaborative filtering with neighborhood contrastive learning.
\end{itemize}

We do not directly compare self-supervised learning~(SSL) baselines for the proposed method is mainly a backbone for CF which is an orthogonal study direction with SSL for a recommendation. In fact in the experiment, we find the performance of JGCF even surpasses SSL methods with LightGCN as the backbone.

\subsubsection{\textbf{Evaluation Metrics.}} We use Recal@$K$ and NDCG@$K$, two widely used metrics for collaborative filtering to evaluate the performance of top-$K$ recommendation. Following~\cite{he2020lightgcn,wu2021self,lin2022improving}, the full-ranking strategy~\cite{10.1145/3394486.3403226,zhao2020revisiting} is adopted, where we rank all the candidate items that the user has not interacted with. We report the values with $K=10,20,50$ to show the performance in different ranges. Higher Recall@$K$ and NDCG@$K$ mean better performance. The metric is averaged for all users in the test set.

\subsubsection{\textbf{Implementation Details.}} We implement the proposed methods and baselines with RecBole~\cite{zhao2021recbole}, a unified open-source framework for developing and reproducing recommendation algorithms. We optimize all the baselines with Adam and carefully choose hyper-parameters according to their suggestions. The batch size is set to 4096 for Gowalla, Yelp, and Alibaba-iFashion and 8192 for Amazon-Books. All the parameters are initialized by Xavier distribution. The embedding size is set to 64 for all methods. Early stopping with the patience of 5 epochs is utilized to prevent overfitting, with Recall@20 as the indicator. For JGCF, a and b are tuned in [-1.0,2.0] with step size 0.5, the learning rate is fixed to 0.001, $\lambda$ is fixed to $1e{-6}$ and the order $K$ of the polynomial is searched in \{1,2,3,4\}. The experiments are based on a 24GB RTX 4090 GPU. 

\subsection{Overall Performance}

\begin{table*}[t]
    \centering
    \Large
    \caption{Main Experimental Results with CF Methods. The best result is \textbf{bolded}, and the runner-up is underlined. \textbf{*} indicates the statistical significance for $p<0.01$ compared to the best baseline. OOM means out-of-memory.}
    \resizebox{.9\linewidth}{!}{
    \begin{tabular}{cc|ccccccccc|cc}
    \toprule
    \multicolumn{2}{c}{\textbf{Setting}} & \multicolumn{9}{c}{\textbf{Baseline Methods}} & \multicolumn{2}{c}{\textbf{Ours}} \\ 
       Dataset  & Metric & BPR & NeuMF &  NGCF & DGCF & LightGCN & GDE & GTN & RGCF  & DirectAU & JGCF & Improv. \\
        \midrule
       \multirow{6}{*}{Gowalla} & Recall@10 & 0.1159 & 0.0975 & 0.1119 &  0.1252 &  0.1382 & 0.1395 & \underline{0.1403} & 0.1335 &
0.1394 & $\textbf{0.1574}^*$ & $ 12.19\%$ \\
       & NDCG@10 & 0.0811 & 0.0664  & 0.0787 & 0.0902 & 0.1003 & 0.1008 & \underline{0.1009} & 0.0905 &  0.0991 &  $\textbf{0.1145}^* $ & $ 13.48\%$ \\
       & Recall@20 & 0.1686 & 0.1470 & 0.1633 & 0.1829 & 0.1983 & 0.1989 & \underline{0.2016} & 0.1934 & 0.2014 & $\textbf{0.2232}^*$ & $ 10.71\%$ \\
       & NDCG@20 & 0.0965 & 0.0808  & 0.0937 & 0.1066 & 0.1175 & 0.1183 & \underline{0.1184} & 0.1081 & 0.1170 &  $\textbf{0.1332}^*$  & $12.50\%$ \\
       & Recall@50 & 0.2715 & 0.2401  & 0.2641 & 0.2877 & 0.3067 & 0.3059 & \underline{0.3132} & 0.3037 & 0.3127 & $\textbf{0.3406}^*$ & $8.75\%$ \\
       & NDCG@50 & 0.1215 & 0.1035  & 0.1182 &  0.1322 & 0.1438 & 0.1448 &\underline{0.1456} & 0.1351  & 0.1442 & $\textbf{0.1619}^*$ &  $11.20\%$ \\
       \midrule
       \multirow{6}{*}{Amazon-Books} & Recall@10 & 0.0477 & 0.0342 & 0.0475 & 0.0565  & 0.0620 & OOM & 0.0588 & \underline{0.0712} & 0.0683 & $\textbf{0.0796}^*$ & 11.80\% \\
       & NDCG@10 & 0.0379 & 0.0266  & 0.0330 & 0.0448 & 0.0506 & OOM & 0.0485 & \underline{0.0568}  & 0.0569 & $\textbf{0.0671}^*$ & 18.13\% \\
       & Recall@20 & 0.0764 & 0.0575  & 0.0760 & 0.0867 & 0.0953 & OOM & 0.0930 & \underline{0.1090} & 0.1053 & $\textbf{0.1191}^*$ & 9.27\% \\
       & NDCG@20 & 0.0474 & 0.0345  & 0.0472 & 0.0551 & 0.0615 & OOM & 0.0597 & \underline{0.0697} & 0.0689 & $\textbf{0.0798}^*$ & 14.49\% \\
       & Recall@50 & 0.1357 & 0.1085 & 0.1362 & 0.1497 & 0.1634 & OOM & 0.1630 & \underline{0.1800} & 0.1760 & $\textbf{0.1949}^*$ & 8.28\% \\
       & NDCG@50 & 0.0647 & 0.0493 & 0.0647 & 0.0735 & 0.0813 & OOM & 0.0800 & \underline{0.0906} & 0.0894 & $\textbf{0.1019}^*$ & 12.47\% \\
       \midrule
       \multirow{6}{*}{Yelp}  & Recall@10 & 0.0452 & 0.0313 & 0.0459 & 0.0527 & 0.0560 & 0.0483 & 0.0603 & \underline{0.0633} & 0.0557 & $\textbf{0.0687}^*$ & 8.53\%  \\
       & NDCG@10 & 0.0355 & 0.0235 & 0.0364 & 0.0419 & 0.0450 & 0.0383 & 0.0483 & \underline{0.0503} & 0.0435 & $\textbf{0.0556}^*$ & 10.54\% \\
       & Recall@20 & 0.0764 & 0.0548  & 0.0778 &  0.0856 & 0.0913 & 0.0808 & 0.0984 & \underline{0.1026} & 0.0907 & $\textbf{0.1105}^*$ & 7.70\% \\
       & NDCG@20 & 0.0460 & 0.0316  & 0.0472 & 0.0528 & 0.0569 & 0.0493 & 0.0611 & \underline{0.0637} & 0.0553 & $\textbf{0.0694}^*$ & 8.95\% \\
       & Recall@50 & 0.1431 & 0.1114 & 0.1461 & 0.1598 & 0.1680 & 0.1514 & 0.1770 & \underline{0.1838} & 0.1665 & $\textbf{0.1953}^*$ & 6.26\% \\
       & NDCG@50 & 0.0653 & 0.0480 & 0.0645 & 0.0743 & 0.0790 & 0.0698 & 0.0839 & \underline{0.0874} & 0.0773 &$\textbf{0.0940}^*$ & 7.55\% \\
       \midrule
       \multirow{6}{*}{Alibaba-iFashion} & Recall@10 &  0.0303 & 0.0182  &  0.0382 & 0.0447 &  \underline{0.0477} & OOM & 0.0406 & 0.0450 & 0.0319 & $\textbf{0.0598}^*$ & 25.37\% \\
        & NDCG@10 & 0.0161 & 0.0092 & 0.0198 &  0.0241 & \underline{0.0255} & OOM & 0.0217 & 0.0249 & 0.0166 & $\textbf{0.0324}^*$ & 27.06\% \\
         & Recall@20 &  0.0467 & 0.0302  & 0.0615 & 0.0677 & \underline{0.0720} & OOM & 0.0625 & 0.0674 & 0.0484 & $\textbf{0.0875}^*$ & 21.53\% \\
         & NDCG@20 & 0.0203 & 0.0123 &  0.0257 & 0.0299 &  \underline{0.0316} & OOM & 0.0272 & 0.0305 & 0.0207 & $\textbf{0.0394}^*$ & 24.68\% \\
         & Recall@50 & 0.0799 & 0.0576  & 0.1081 & 0.1120  & \underline{0.1187} & OOM & 0.1070 & 0.1110 & 0.0768 & $\textbf{0.1379}^*$ & 16.18\% \\
         & NDCG@50 &  0.0269 & 0.0177 & 0.0349 & 0.0387 & \underline{0.0409} & OOM & 0.0361 & 0.0392 & 0.0264 & $\textbf{0.0494}^*$ & 20.78\% \\ 
    \bottomrule
    \end{tabular}}
    % \begin{tablenotes}
    % \normalsize
    % \flushleft
    
    % \end{tablenotes}
    \label{tab:main_exp}
\end{table*}

 The experimental results for four public datasets are shown in Table~\ref{tab:main_exp}. From the table, we have the following observations:

1) GNN-based methods are relatively better than conventional matrix factorization methods. The main reason is due to the nature of GNN-based models to incorporate high-order information in the use-item bipartite graph, while MF methods only use the first-order neighbors to conduct model training. Besides, only using propagation~(LightGCN) can perform better than complex or heavy model structures~(NGCF and DGCF). Complex designs or heavy structures are easy to overfit. We also find that recently denoising GNN methods~(GTN and RGCF) achieve good performance on most datasets. They can remove the influence of noisy interactions and preserve the most important links. The difference is that GTN uses sparsity optimization based on graph trend filtering while RGCF directly manipulates the graph structure. Despite the effectiveness, in Figure~\ref{fig:time_cost} we find RGCF takes more time than GTN to train. Also, we find decomposition-based method GDE can not scale for larger datasets like Amazon-Books and Alibaba-iFashion.

2) JGCF is consistently better than all the baselines on all the public datasets, showing the effectiveness of the improvement of the graph filtering strategy. JGCF tends to suppress the middle-frequency signals which do not contain too much information about user interests. On the contrary, it will emphasize the low-frequency signals which contain information about similar users and similar items. Moreover, it will also emphasize the high-frequency signal which contains information about the difference between similar users and similar items, giving the model the ability to distinguish fine-grained differences. It may alleviate the over-smoothing issue.

3) One more interesting discovery is that JGCF tends to perform better on sparser graphs. From Table~\ref{tab:main_exp}, we find that compared to the relative improvement on Yelp, JGCF has a larger margin of increases on Gowalla, Amazon-books, and Alibaba-iFashion, which are sparser than Yelp. Moreover, JGCF has the most improvement on Alibaba-iFashion, which is the sparsest dataset with a sparsity rate of 99.99\%. This is more consistent with real-world E-commerce scenarios, as user interactions follow the power-low distribution and most users have few interactions. The results show the potential of JGCF to handle cold-start users.

% Direct Au yelp
 % {'recall@10': 0.0698, 'recall@20': 0.1126, 'recall@50': 0.1953, 'mrr@10': 0.0951, 'mrr@20': 0.1024, 'mrr@50': 0.1075, 'ndcg@10': 0.0561, 'ndcg@20': 0.0704, 'ndcg@50': 0.0942, 'hit@10': 0.2354, 'hit@20': 0.342, 'hit@50': 0.5052, 'precision@10': 0.0289, 'precision@20': 0.0235, 'precision@50': 0.0164}

\subsection{Ablation Study}
\begin{table}[t]
\Large
    \centering
    \caption{Ablation Study of JGCF.}
    \resizebox{.95\linewidth}{!}{
    \begin{tabular}{c|cc|cc|cc}
    \toprule
       \multirow{2}{*}{Variants} & \multicolumn{2}{c}{Gowalla} & \multicolumn{2}{c}{Amazon-Books} & \multicolumn{2}{c}{Yelp} \\
        & R@20 & N@20 & R@20 & N@20 & R@20 & N@20 \\
       \midrule
        JGCF & \textbf{0.2232} & \textbf{0.1332} & \textbf{0.1191} & \textbf{0.0798} & \textbf{0.1105} & \textbf{0.0694} \\
        \midrule
        JGCF w/o BS & 0.2112
 & 0.1242 & 0.1031 & 0.0674 & 0.0856 &  0.0528 \\
        JGCF w/o BP & 0.2195 & 0.1299 & 0.1163 & 0.0770 & 0.1067 & 0.0664 \\
        JGCF w/o act & 0.2224 & 0.1322 & 0.1184 & 0.0782 & 0.1091 & 0.0687 \\
        \bottomrule 
    \end{tabular}}
    \label{tab:ablation}
\end{table}
In this section, we study the performance of several variants of JGCF. The different variants are 1) without frequency decomposition and only use band-stop components~(JGCF w/o BP), 2) without band-stop components and only use the band-pass components~(JGCF w/o BS). The results are shown in Table~\ref{tab:ablation}. From the table, we find that both of the components are effective and JGCF without any of the components will have a performance degeneration. However, the degeneration degree is different. Compared with the band-pass component, the band-stop component contributes more to the final recommendation accuracy, showing the effectiveness of suppressing the middle-frequency signal and emphasizing the low and high-frequency components. 
% \subsection{Sparsity Study}

\subsection{Hyper-parameter Study}

\subsubsection{\textbf{Effect of the polynomial order $K$}}
\begin{figure}[t]
    \centering
    \subfigure[Gowalla]{\includegraphics[width=.45\linewidth]{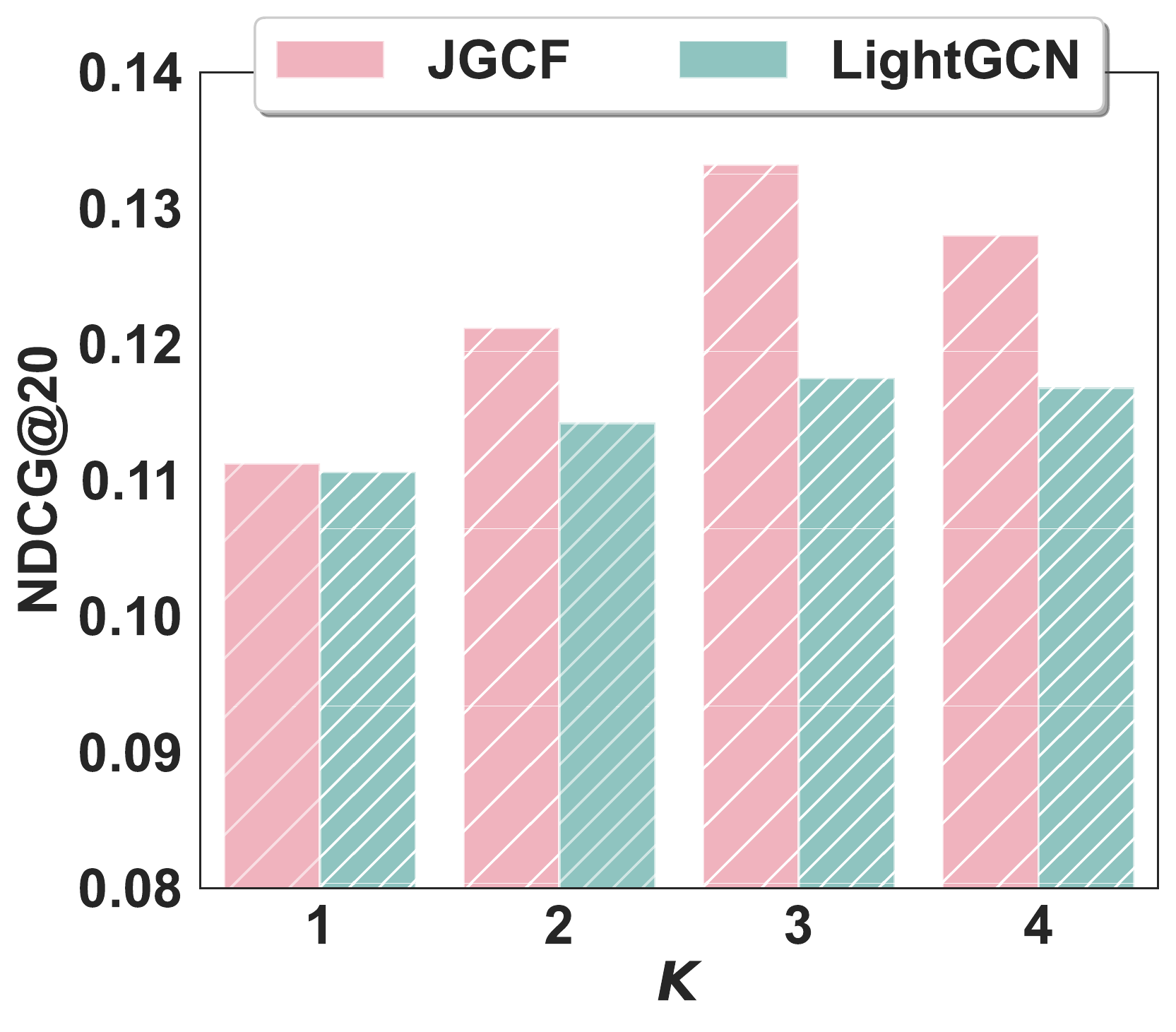}}
    \subfigure[Amazon-Books]{\includegraphics[width=.45\linewidth]{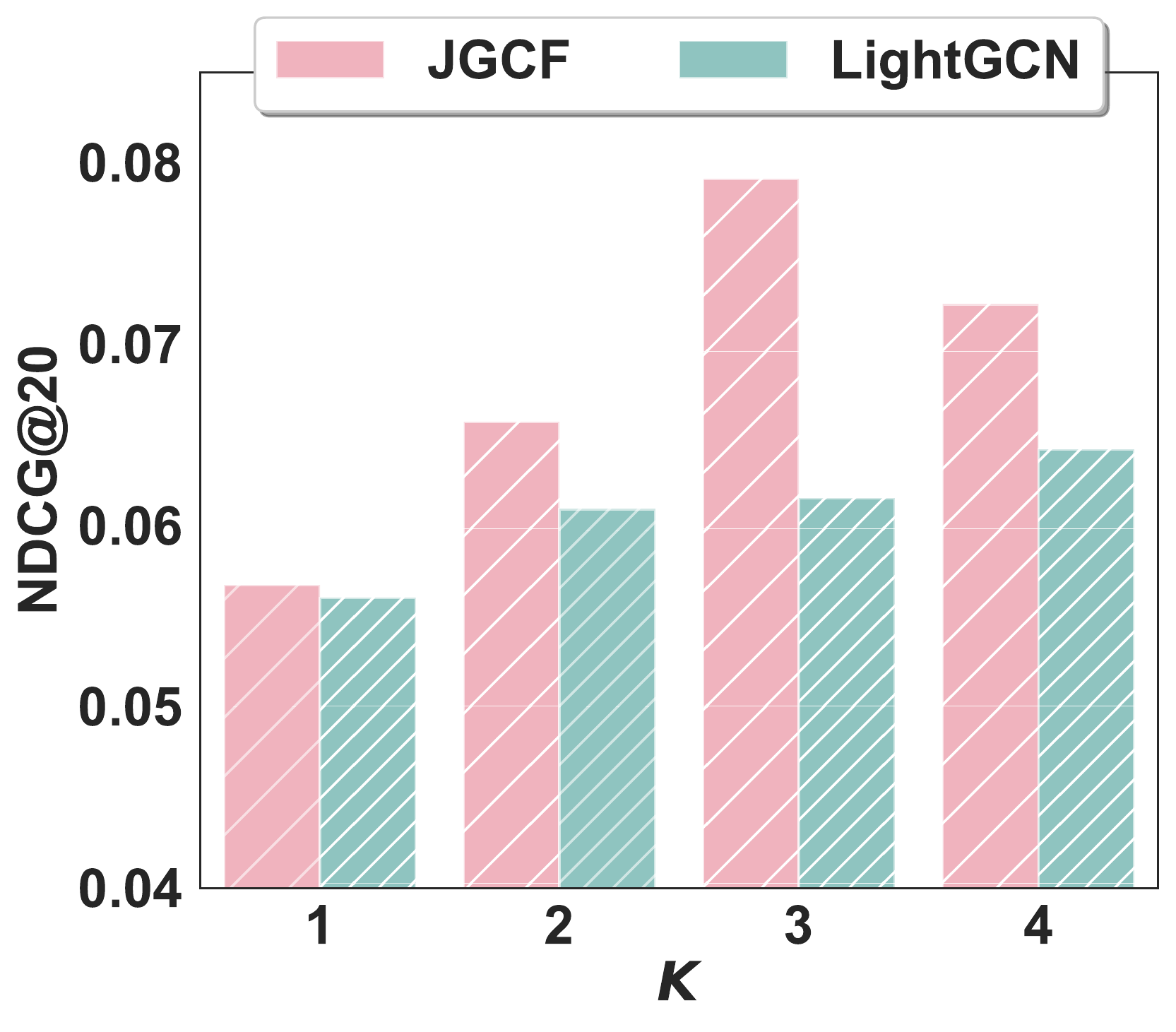}}
    \subfigure[Yelp]{\includegraphics[width=.45\linewidth]{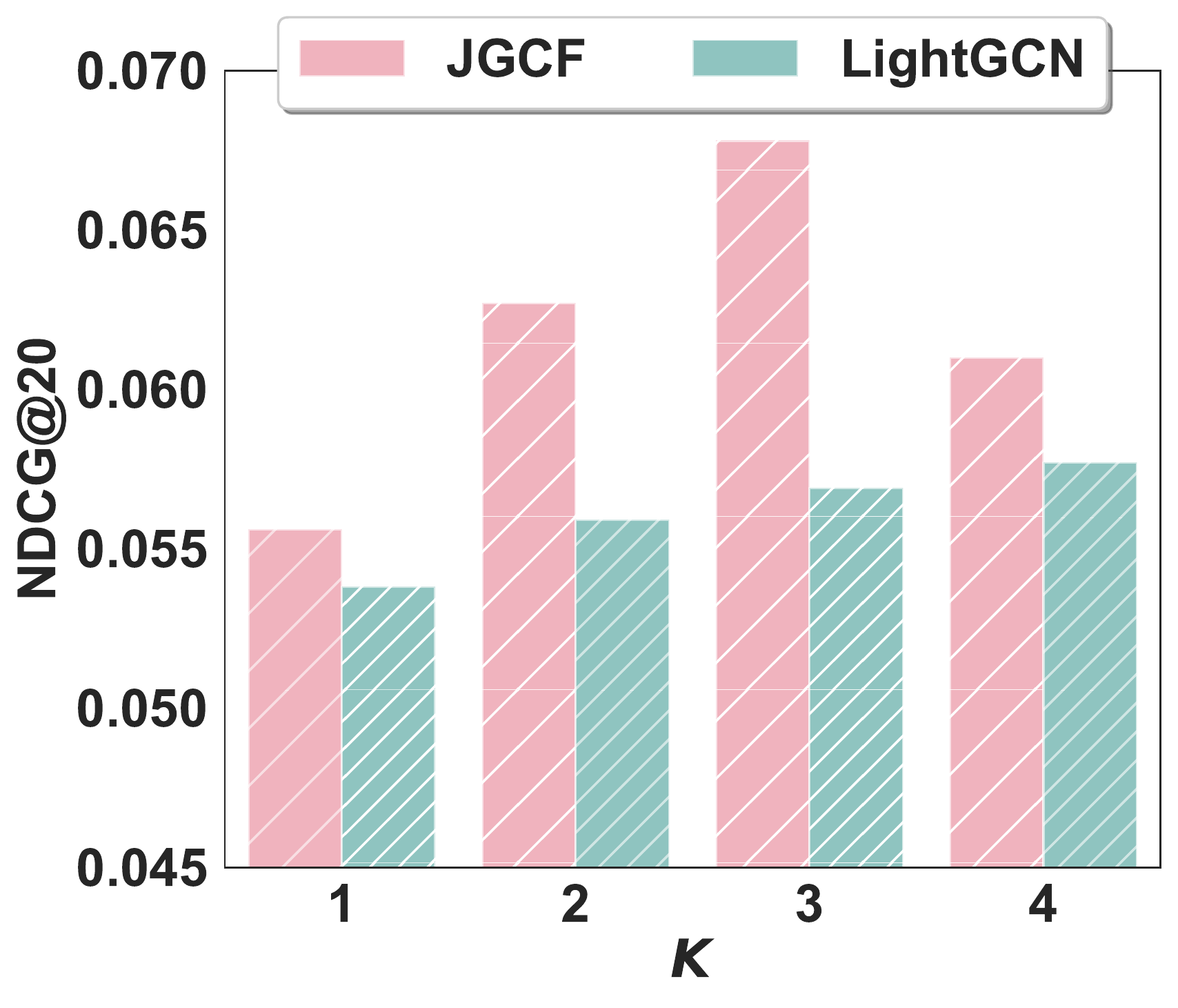}}
    \subfigure[Alibaba-iFashion]{\includegraphics[width=.45\linewidth]{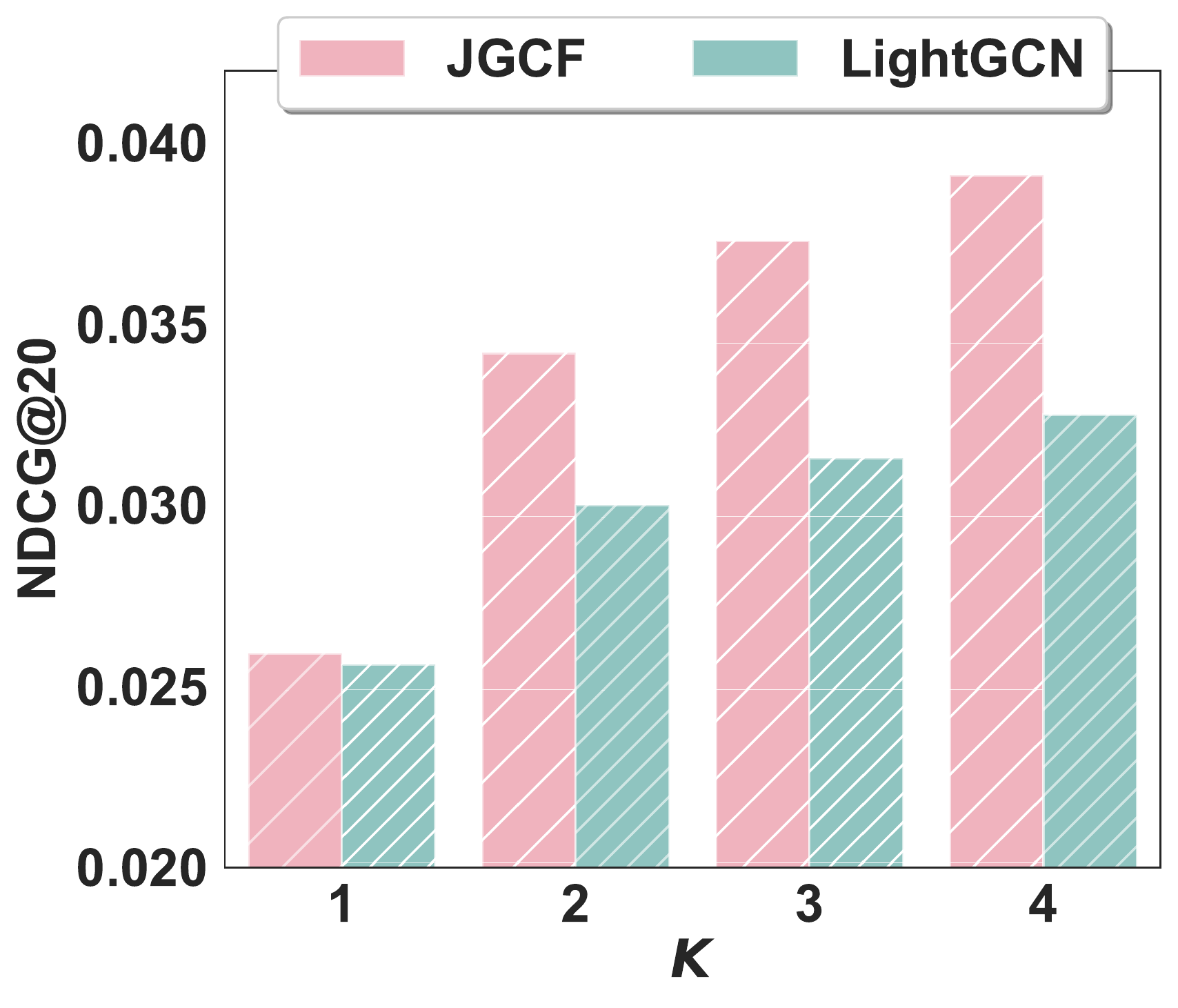}}
    \caption{Analysis of the order $K$.}
    \label{fig:hyper_K}
\vspace{-0.2cm}
\end{figure}
To study the effect of the order of the polynomials, we keep $a=b=1.0$, $\alpha=0.1$ and vary $K$ from 1 to 4. The results are shown in Figure~\ref{fig:hyper_K}. We find that in general larger $K$ leads to better performance. It is found both for JGCF and LightGCN and is consistent  with previous studies~\cite{he2020lightgcn,lin2022improving}. Besides, with $K$ increases, the performance of JGCF increases faster than LightGCN, showing the supreme of the orthogonality of Jacobi polynomials and its effectiveness in compressing irrelevant signals and strengthening important ones. In fact, approximating the graph filter using orthogonal polynomials will make it converge faster, which is important in industry scenarios as many real-world applications need to update the model parameters in a limited time. 

\subsubsection{\textbf{Impact of the parameters a and b}}
\begin{figure}[t]
    \centering
    \subfigure{\includegraphics[width=.44\linewidth]{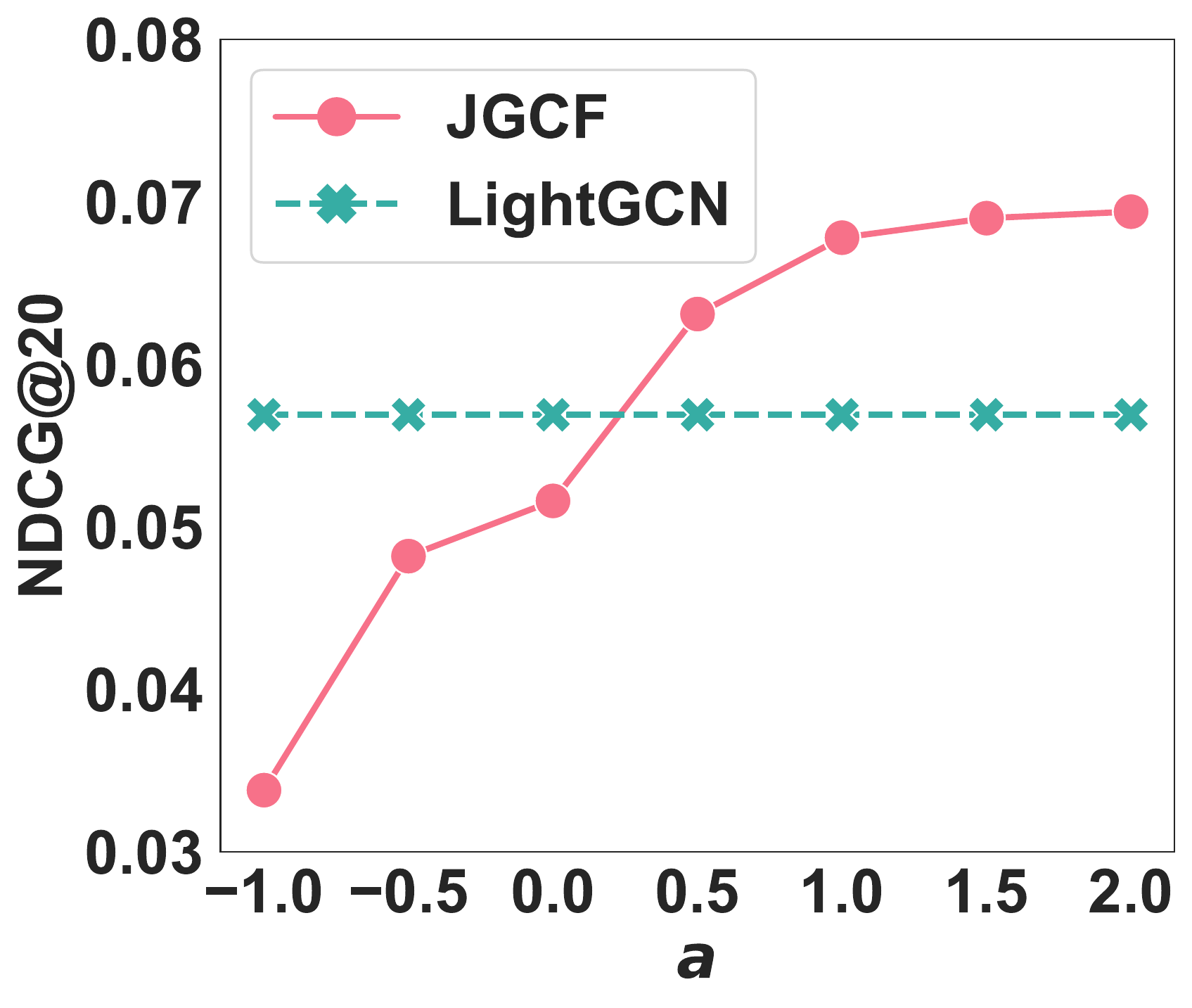}}
    \subfigure{\includegraphics[width=.46\linewidth]{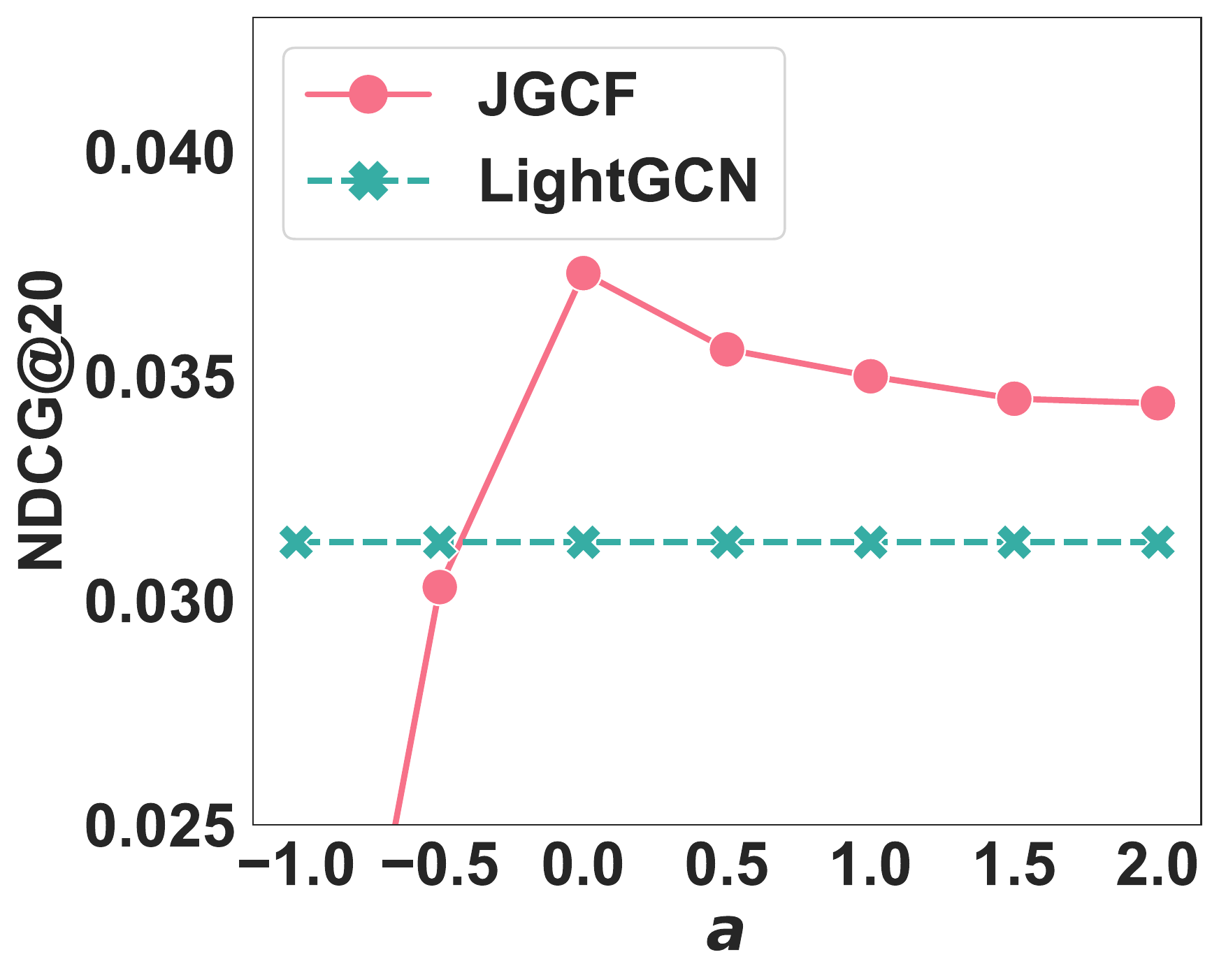}}
    \subfigure{\includegraphics[width=.44\linewidth]{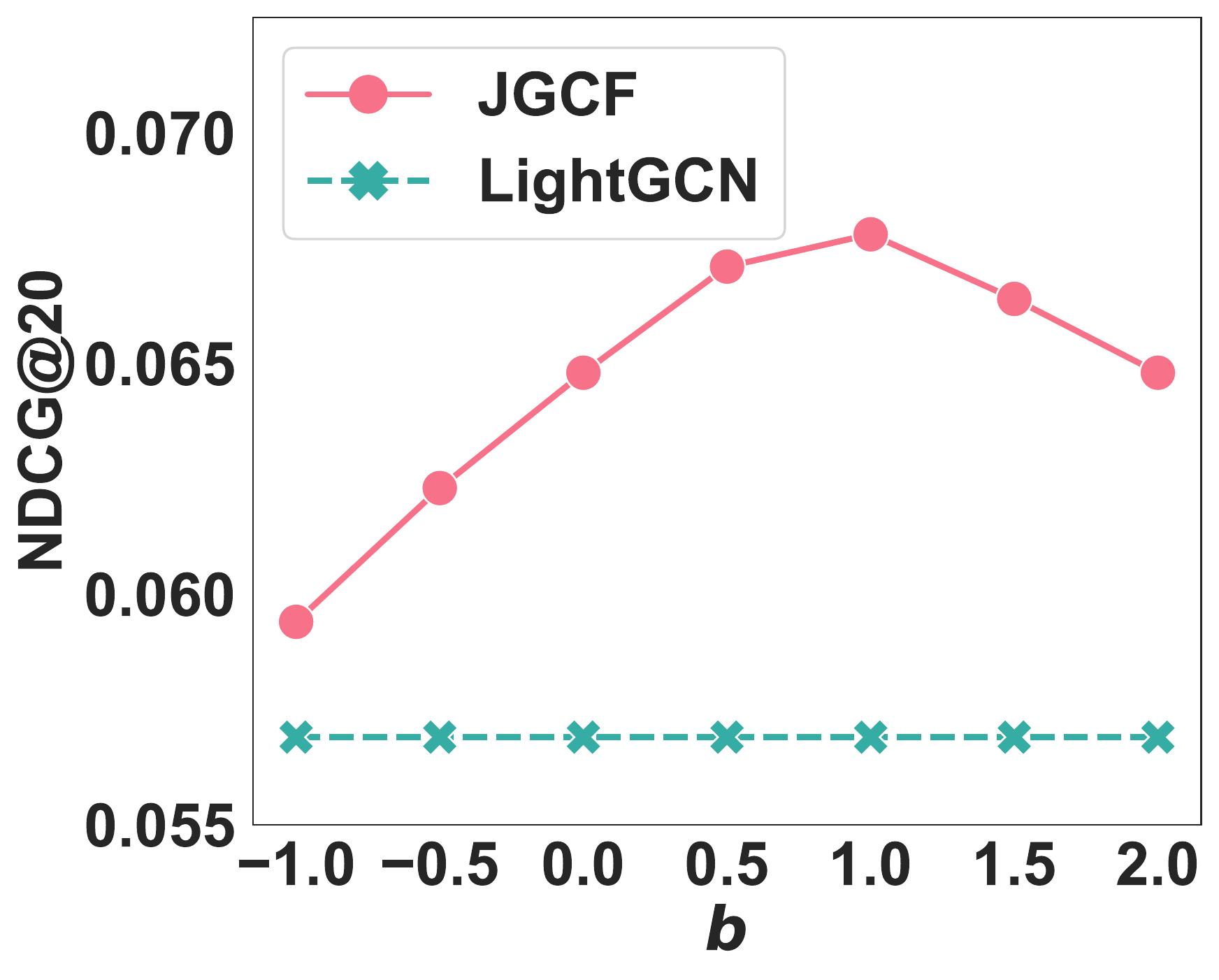}}
    \subfigure{\includegraphics[width=.46\linewidth]{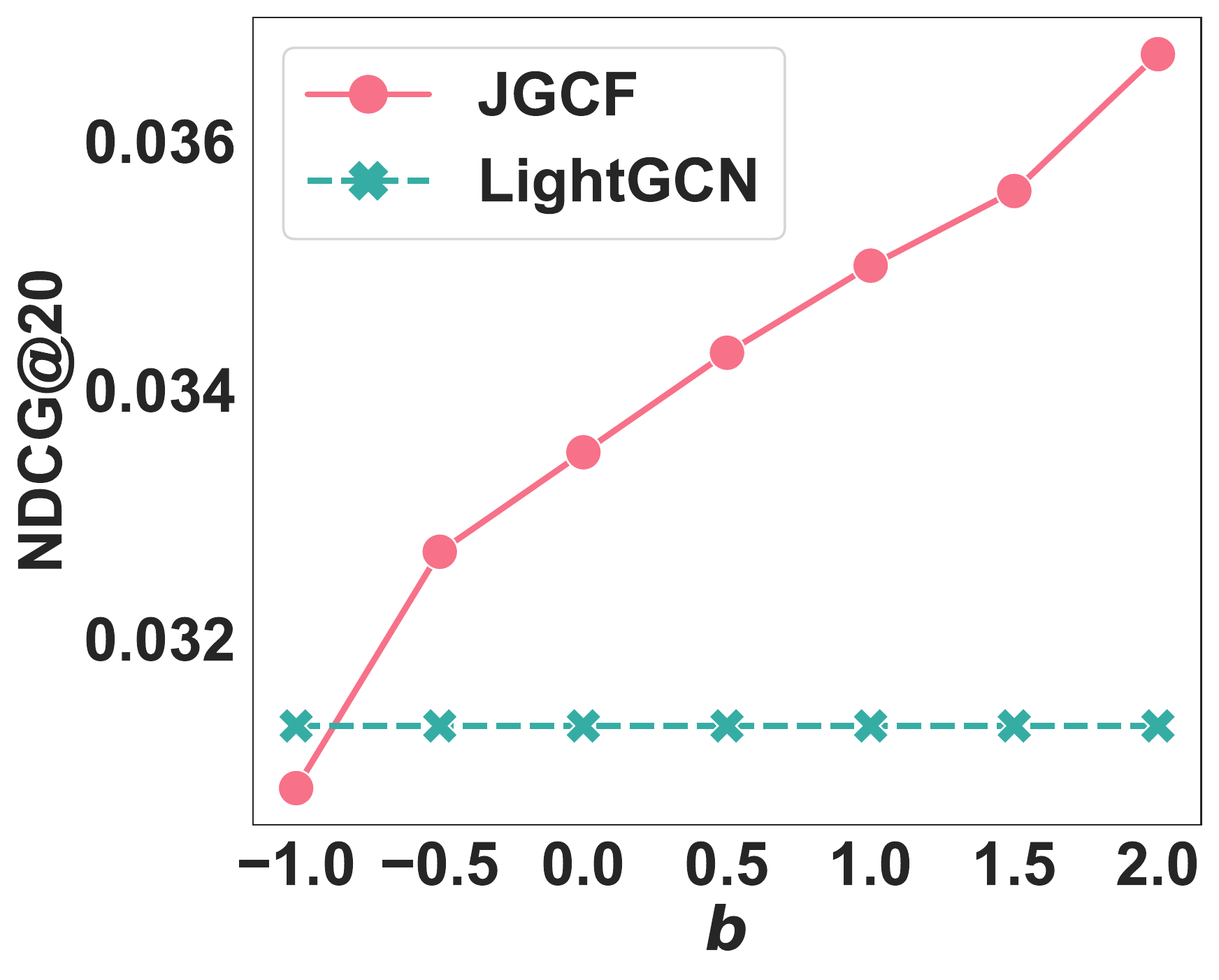}}
    \subfigure{\includegraphics[width=.44\linewidth]{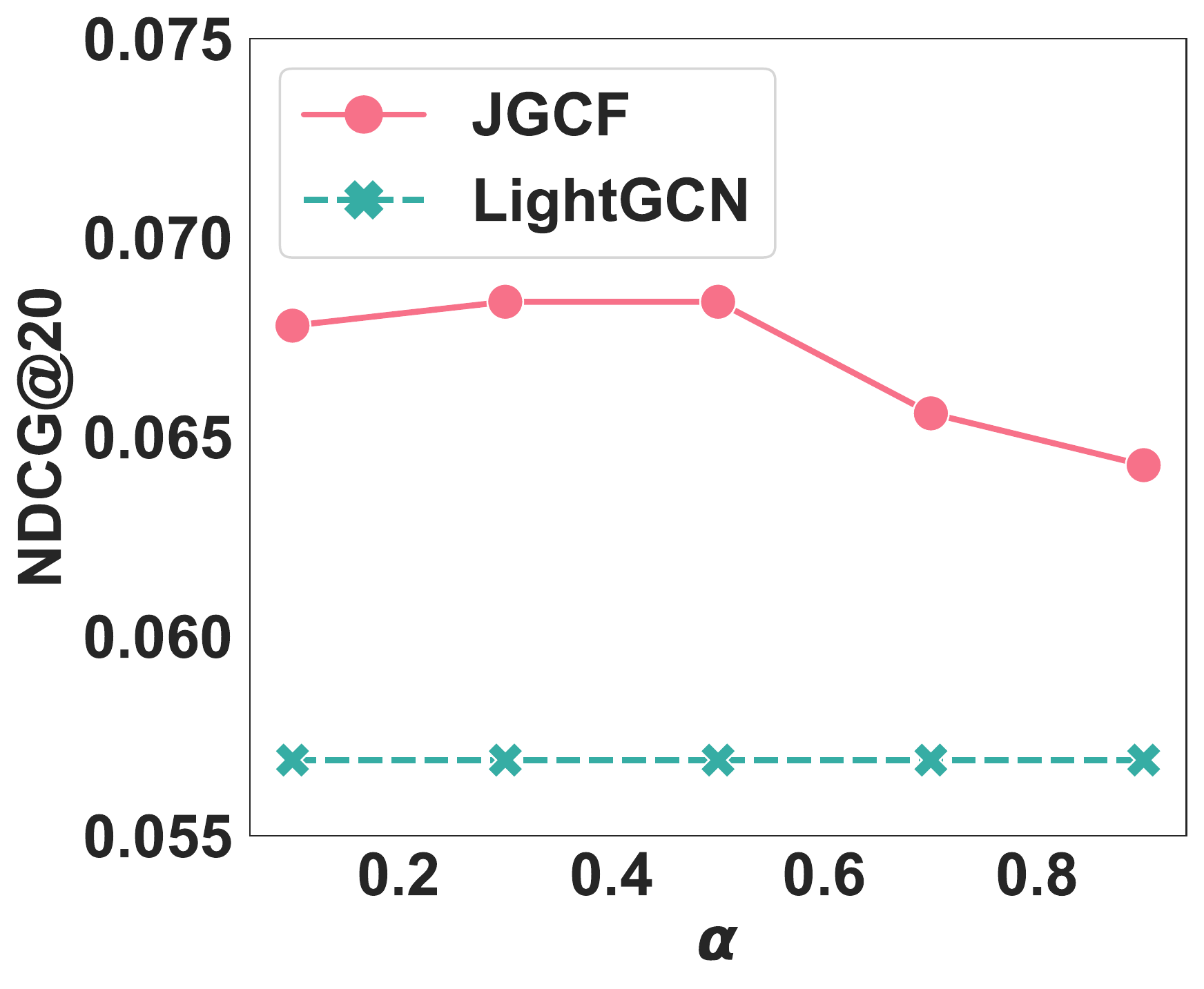}}
    \subfigure{\includegraphics[width=.46\linewidth]{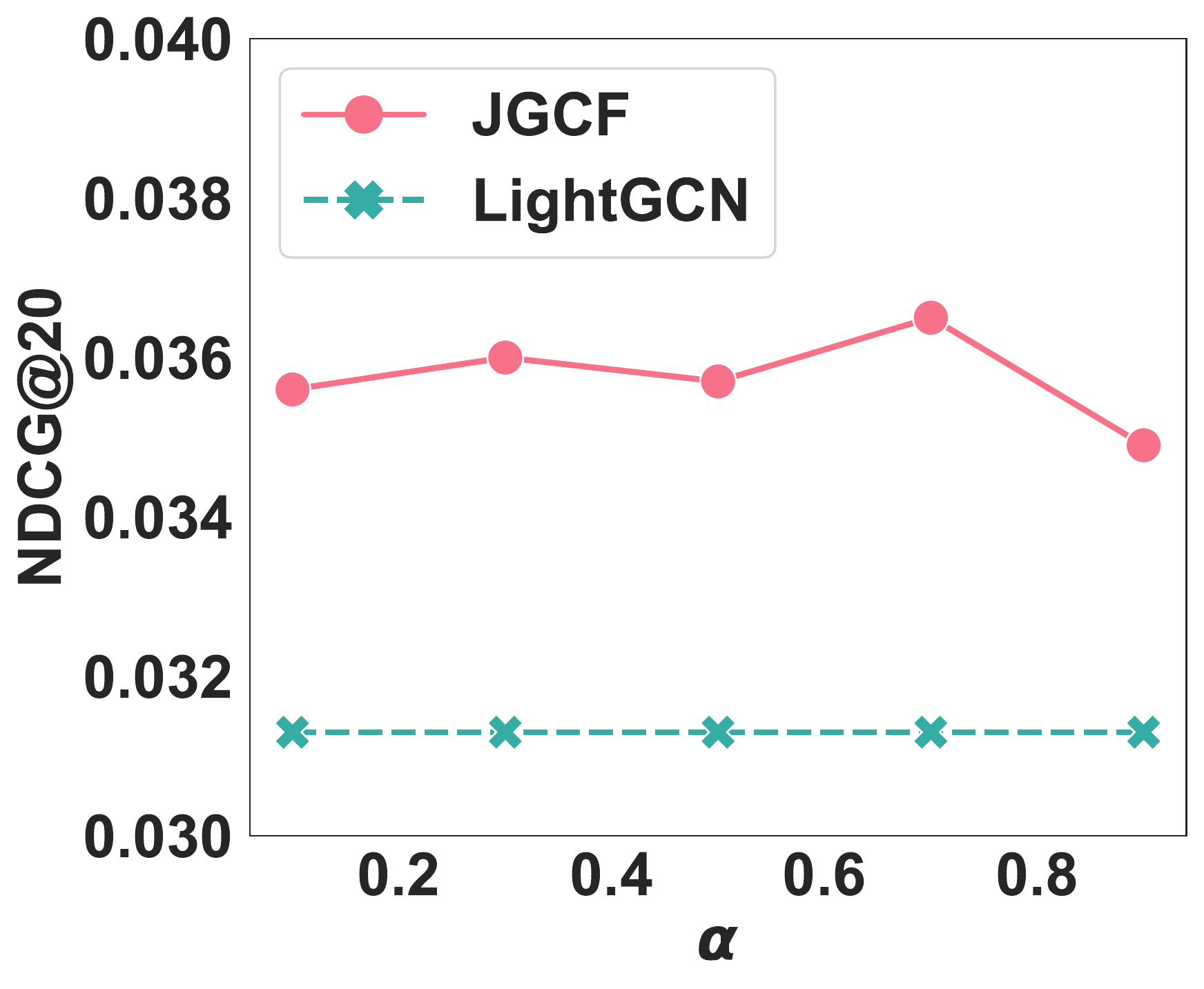}}
    \caption{Analysis of $a$, $b$ and $\mathbf{\alpha}$. The results for Yelp and Alibaba-iFashion are shown in the left and the right column.}
    \label{fig:hyper_ab}
\end{figure}\
The result for the parameters $a$ and $b$ is shown in Figure~\ref{fig:hyper_ab}. $a$ and $b$ control the behavior of the graph filter. The relative values of $a$ and $b$ control the influence of low and high-frequency signals. For fair comparisons, we fix $a=1.0$, $\alpha=0.1$ when studying $b$ and $b=1.0$, $\alpha=0.1$ when studying $a$. We observe that $b$ is more important in influencing the performance of JGCF, as low-frequency signal contains much information about similar users and items. And we should not  keep $b$ too large as emphasizing low-frequency signals too much will cause over-smoothing in some datasets like Yelp. For $a$, we find that in Yelp, larger $a$ leads to better results. It is because high-frequency signal makes the models learns the difference among similar nodes, which brings more knowledge. Moreover, it may bring randomness and leaves less chance of over-fitting.

\subsubsection{\textbf{Effect of $\mathbf{\alpha}$}}

$\alpha$ mainly controls the influence of the band-pass part. $\alpha=0$ means the model goes back to a band-stop filter and $\alpha=1$ means the model will have more weight on the middle-frequency components. We find that including too much middle-frequency signal will degenerate the overall performance. However, if we restrict $\alpha$ in a proper range, the model also shows improved performance. The proper range is varied for different datasets. For example, for Yelp $\alpha$ need to be smaller than 0.5 while for Alibaba-iFashion the proper choice for $\alpha$ is around 0.7.
\begin{table}[t]
    \centering
    \caption{Performance as the backbone.}
    \resizebox{.8\linewidth}{!}{
    \begin{tabular}{c|cc|cc}
    \toprule
   Metrics & \thead{SGL w/ \\ LightGCN} & \thead{SGL w/ \\ JGCF} & \thead{NCL w/ \\ LightGCN} & \thead{NCL w/ \\ JGCF} \\
    \midrule
       Recall@10  & 0.1403 & \textbf{0.1533} & 0.1528 & \textbf{0.1552} \\
       NDCG@10  & 0.1028 & \textbf{0.1119} & 0.1102 & \textbf{0.1130} \\
       Recall@20 & 0.1998 & \textbf{0.2183} & 0.2146 & \textbf{0.2214} \\
       NDCG@20 & 0.1199 & \textbf{0.1304} & 0.1280 & \textbf{0.1318} \\
       Recall@50 & 0.3097 &  \textbf{0.3343} & 0.3228 &  \textbf{0.3380} \\
       NDCG@50 & 0.1467 & \textbf{0.1588} & 0.1547 & \textbf{0.1602} \\
    \bottomrule
    \end{tabular}}
    \label{tab:backbone}
\end{table}

\subsection{Further Analysis}
\subsubsection{\textbf{Performance as Backbone}}

We replace the LightGCN backbone in SGL and NCL to further verify the generalization ability of JGCF for self-supervised learning. The results are shown in Table~\ref{tab:backbone}. Surprisingly, we observe that by replacing LightGCN with JGCF, the performance will have a large increase, indicating that suppressing the middle signal while at the same time emphasizing the low-frequency and high-frequency components is also more suitable for contrastive learning. 
% Due to space limitations, we only report the result for Gowalla while we find a similar trend in the other four datasets.
\begin{figure}[t]
    \centering
    \includegraphics[width=.9\linewidth]{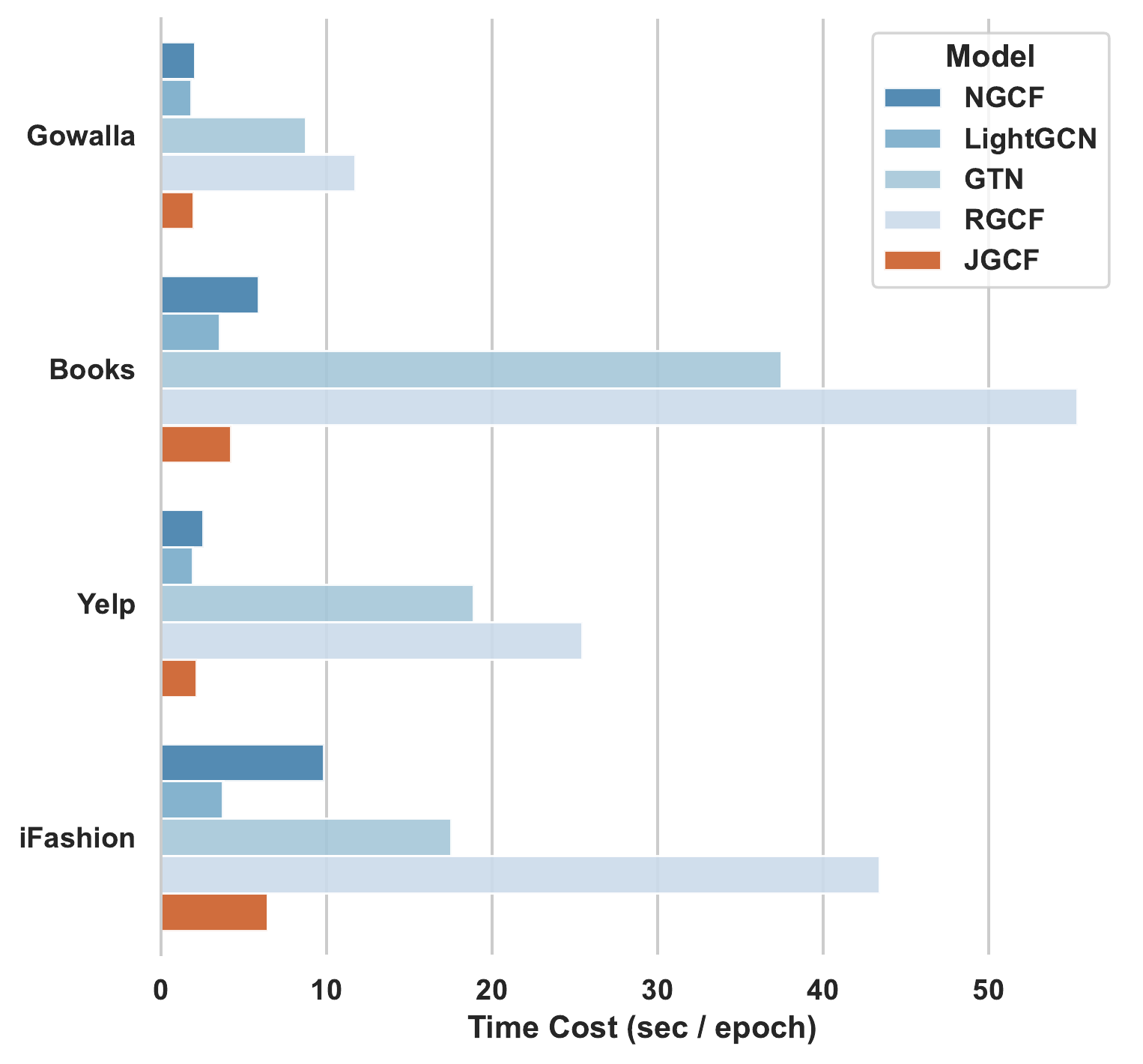}
    \caption{Time cost per training epoch.}
    \label{fig:time_cost}
\end{figure}
\subsubsection{\textbf{Comparison with Other Polynomial bases}}
In this section, we compare Jacobi polynomial basis with other widely used ones. The results are shown in Table~\ref{tab:poly}. We find orthogonal polynomial bases is consistently better than the monomial bases for CF, showing the effectiveness of the orthogonality of polynomials in achieving lower approximation error. Besides, among those orthogonal polynomials, the Jacobi polynomial with the best parameters consistently works better than Chebyshev and Legendre, showing that the fixed parameter polynomials may not be the best choice. We also compare with BernNet to show that too complex adaptive filters may overfit thus affecting the performance.
\subsubsection{\textbf{Efficiency Comparison}} We compare the training time per epoch of some baselines and JGCF on Gowalla. All the methods use the same batch size, model depth, and embedding size for fair comparisons. The results are shown in Table~\ref{fig:time_cost}. From the table, we find that JGCF is almost as efficient as LightGCN, which takes less time than NGCF on the datasets. It is due to the reason that JGCF and LightGCN have the same number of parameters which is smaller than NGCF. Moreover, the two SOTA methods GTN and RGCF take more than tens of the training time per epoch of JGCF on the four datasets for their heavier designs. We do not include the results of DGCF for it takes too much time to train for an epoch. In a word, JGCF is also an efficient method for collaborative filtering. 
\begin{table}[t]
\Large
    \centering
    \caption{Comparison of popular polynomial basis.}
    \resizebox{.85\linewidth}{!}{
    \begin{tabular}{c|cccc}
    \toprule
       \multirow{2}{*}{Polynomials} & \multicolumn{2}{c}{Yelp} & \multicolumn{2}{c}{Alibaba-iFashion} \\
       & Recall@20 & NDCG@20 & Recall@20 & NDCG@20 \\
    \midrule
       Monomial  & 0.0926 & 0.0573 & 0.0701 & 0.0309 \\
       Chebyshev & 0.1015 & 0.0637 & 0.0818 & 0.0364 \\
       Legendre  & 0.0966 & 0.0598 & 0.0809 & 0.0362 \\
       BernNet   & 0.0808 & 0.0493 & 0.0562 & 0.0244 \\
       Jacobi    & $\textbf{0.1067}$ &  $\textbf{0.0664}$ & $\textbf{0.0875}$ & $\textbf{0.0389}$ \\
    \bottomrule
    \end{tabular}}
    \label{tab:poly}
\end{table}

\section{Related Works}

\subsection{Spectral Graph Neural Network}

Spectral graph neural networks~\cite{pmlr-v162-wang22am,xu2018graph,defferrard2016convolutional,dong2021adagnn,he2022chebnetii,he2021bernnet,kipf2017semi,chien2020adaptive,bianchi2021graph,zhu2020simple,huang2023robust} are designed initially with graph signal processing~\cite{ortega2018graph,dong2020graph,gavili2017shift}, which aim to design signal filters on the graph spectral domain to facilitate the downstream task learning. Defferrard et al.~\cite{defferrard2016convolutional} uses Chebyshev polynomials to fastly approximate the graph convolution operation. Kipf et al.~\cite{kipf2017semi} simplifies the graph convolution by using the first-order Chebyshev polynomial. Xu et al.~\cite{xu2018graph} use graph wavelet transformation to design convolution operations. He et al.~\cite{he2021bernnet} propose an adaptive filter learning method based on Bernstein polynomials. Wang and Zhang~\cite{pmlr-v162-wang22am} study the expressive power of spectral graph neural networks. Compared with spatial GNN, spectral GNN enjoys better theoretical background and explainability. It is also more efficient than attention-based GNN~\cite{velickovic2018graph} thus it is well-suited for large graph datasets.

\subsection{Graph Collaborative Filtering}

Graph nerual networks~\cite{bi2023predicting,bi2022company,bi2023mm,zhang2023continual,guo2023homophily} have been a popular choice for recommender systems~\cite{guo2022learning,guo2022evolutionary,zhang2023efficiently}. Graph collaborative filtering~(GCF) aims to conduct collaborative filtering with graph processing. It provides a general framework to incorporate high-order information in the user-item graph to boost the recommendation performance. Wang et al.~\cite{wang2019neural} first propose the graph collaborative filtering framework to make recommendations. Wang et al.~\cite{wang2020disentangled} further adopt disentangled propagation to enhance the recommendation. Fan et al.~\cite{fan2022graph} uses graph trend filtering to remove the effect of noisy edges and enable negative weight propagation. Tian et al.~\cite{tian2022learning} design a structure learning approach to remove noisy interactions while preserving diversity by adding new edges. Peng et al.~\cite{10.1145/3477495.3532014} find that low and high-frequency signals are important for GCF.  Despite the design of the model structure, another line of work focuses on improving GCF with self-supervised learning. Wu et al.~\cite{wu2021self} propose the self-supervised learning pipeline for GCF. Lin et al.~\cite{lin2022improving} propose neighborhood contrastive learning to improve GCF. Yu et al.~\cite{yu2022graph} propose a simple contastive method to improve GCF. Many works also focus on designing efficient GCF methods. He et al.~\cite{he2020lightgcn} study the redundant component of GCN and design an appropriate model structure that bypasses NGCF. Mao et al.~\cite{10.1145/3459637.3482291} further simplify the propagation of LightGCN with optimization constraints. Shen et al.~\cite{10.1145/3459637.3482264} and Peng et al.~\cite{10.1145/3511808.3557462} designs SVD-based methods that do not need training. However, the SVD-base methods need to decompose a large adjacency matrix to find top-K eigenvalues and are not scalable.
\section{Conclusion}

In this work, we study graph collaborative filtering from the spectral transformation view and find the characteristics a graph filter should have to make an effective recommendation. We further design JGCF, an efficient and effective method based on Jacobi polynomials and frequency decomposition. Extensive experiments demonstrate the effectiveness of JGCF towards state-of-the-art GCF methods. In the future, we will make JGCF able to process dynamic user-item graphs and enable side information. 

\balance
\bibliography{sample-base}
\bibliographystyle{ACM-Reference-Format}

\nobalance
\appendix

\section{Supplementary Materials}

\subsection{Notations}
We summarize the key notations and their definitions used in this paper in Table~\ref{tab:notation}.
\begin{table}[h]
    \centering
    \caption{Notations}
    \begin{tabular}{c|c}
    \toprule
       Notation  & Definition \\
    \midrule 
       $\mathcal{U}$ & The set for all users. \\
       $\mathcal{I}$ & The set for all items. \\
       $\mathcal{R}$ & The set for all user-item interactions. \\
       $\mathbf{R}$ & The rating matrix of user-item interactions. \\
       $\mathcal{G}$ & User-item bipartite graph. \\
       $||\cdot||_F$ & The Frobenius norm. \\
       $\mathbf{A}$ & The original adjacency matrix of $\mathcal{G}$ \\
       $\hat{\mathbf{A}}$ & The normalized adjacency matrix of $\mathcal{G}$\\
       $\mathbf{D}$ & The diagonals are the degrees of nodes in $\mathcal{G}$ \\
       $\mathbf{U}$ & eigenvectors of $\mathbf{A}$ \\
       $\mathbf{\Lambda}$ & Its diagonals are eigenvalues of $\mathbf{A}$ \\
        $\mathbf{L}$ & Laplacian matrix of $\mathcal{G}$ \\
        $\hat{\mathbf{L}}$ & Normalized Laplacian matrix of $\mathcal{G}$ \\
        $\tilde{\mathbf{U}}$ & eigenvectors of $\mathbf{L}$ \\
       $\tilde{\mathbf{\Lambda}}$ & Its diagonals are eigenvalues of $\mathbf{L}$. \\
       $g_\theta(\cdot)$ & Filter function on the spectrum. \\
       $\mathbf{P}_k$ & Some polynomials with order $k$ \\
       $\mathbf{P}^{(a,b)}_k$ & Jacobi polynomial of order $k$ with $a$ and $b$ \\
       $\mathbf{E}^{(k)}$ & The embedding of order $k$ \\
       $\mathbf{E}$ & Final output embedding. \\
       $\text{tanh}(\cdot)$  & Tanh function. \\
       $\hat{y}_{ui}$ & Prediction score for user $u$ and item $i$ \\
    \bottomrule
    \end{tabular}
    \label{tab:notation}
\end{table}

\subsection{Implementation and Baselines.}
The code link for JGCF is~\url{https://github.com/SpaceLearner/JGCF}, and the links to the baselines used are listed as follows.
\begin{itemize}
\item\textbf{BPR.} \url{https://recbole.io/docs/user_guide/model/general/bpr.html}
\item\textbf{NeuMF.} \url{https://recbole.io/docs/user_guide/model/general/neumf.html}
    \item\textbf{NGCF.} \url{https://recbole.io/docs/user_guide/model/general/ngcf.html}
    \item\textbf{DGCF.} \url{https://recbole.io/docs/user_guide/model/general/dgcf.html}
    \item\textbf{LightGCN.} \url{https://recbole.io/docs/user_guide/model/general/lightgcn.html}
    \item\textbf{GDE.} \url{https://github.com/tanatosuu/GDE}
    \item\textbf{GTN.} \url{https://github.com/wenqifan03/GTN-SIGIR2022}
    \item\textbf{RGCF.} \url{https://github.com/ChangxinTian/RGCF}
    \item\textbf{DirectAU.} \url{https://github.com/THUwangcy/DirectAU}
    \item\textbf{SGL.} \url{https://recbole.io/docs/user_guide/model/general/sgl.html}
    \item\textbf{NCL.}\url{https://recbole.io/docs/user_guide/model/general/ncl.html}
\end{itemize} 

\subsection{Training Algorithm of JGCF}
The pseudocode for the training algorithm of JGCF is shown in Algorithm~\ref{alg:jgcf} to facilitate understanding.
\begin{algorithm}
\renewcommand{\algorithmicrequire}{\textbf{Input:}}
\renewcommand{\algorithmicensure}{\textbf{Output:}}
\renewcommand{\algorithmicwhile}{\textbf{function}}
\caption{Training Algorithm of JGCF.}
\label{alg:jgcf}
    \begin{algorithmic}[1]
   \REQUIRE user-item interactions $\mathcal{R}$, normalized user-item interaction bipartite graph adjacency matrix $\hat{\mathbf{A}}$, weight of Jacobi polynomials $a$ and $b$, coefficient $\alpha$, polynomial orders $K$.
   \ENSURE trained embeddings $\mathbf{E}^{(0)}$.
   \STATE Randomly initializes parameters $\mathbf{E}^{(0)}$.
   \FOR{mini-batch with $n$ user-item pairs in $\mathcal{R}$}
   \FOR{$k$ in $\{1,...,K\}$}
    \IF{k=1}
    \STATE \begin{equation*}
        \mathbf{E}^{(1)}=\frac{a-b}{2}+\frac{a+b+2}{2}\hat{\textbf{A}}\mathbf{E}^{(0)}
    \end{equation*}
    \ELSE
    \STATE \begin{equation*}
      \begin{split}
\mathbf{E}^{(k)}=&\theta_k\hat{\textbf{A}}\mathbf{P}_{k-1}^{a,b}(\hat{\mathbf{A}})\mathbf{E}^{(0)}+\theta_k'\mathbf{P}_{k-1}^{a,b}\mathbf{E}^{(0)}\\ &-\theta''_k\mathbf{P}_{k-2}^{a,b}(\hat{\textbf{A}})\mathbf{E}^{(0)}
      \end{split} 
   \end{equation*} 
    \ENDIF
   \ENDFOR
   \STATE $\mathbf{E}^{(K)}_{\text{band-stop}}=\frac{1}{K+1}\sum_{k=0}^{K}\mathbf{P}_k^{a,b}(\mathbf{A})\mathbf{E}^{(0)}$
    \STATE $\mathbf{E}_{\text{band-pass}}^{(K)}=\text{tanh}\left (\left(\alpha\mathbf{I}-\frac{1}{K+1}\sum_{k=0}^{K}\mathbf{P}_k^{a,b}(\mathbf{A})\right )\mathbf{E}^{(0)}\right )$
    \STATE $\mathbf{E}=[\mathbf{E}_{\text{band-stop}}^{(K)};\mathbf{E}_{\text{band-pass}}^{(K)}]$
    \STATE $\mathcal{L}_{\text{BPR}}=-\sum_{u\in\mathcal{U}}\sum_{i\in\mathcal{N}_u}\sum_{j\notin\mathcal{N}_{u}}\text{ln}\sigma(\hat{y}_{ui}-\hat{y}_{uj})+\lambda ||\mathbf{E}^{(0)}||^2$
    \STATE Update the parameters $\mathbf{E}^{(0)}$ via gradient descent.
   \ENDFOR
    \end{algorithmic}
\end{algorithm}

\subsection{Commonly Used Polynomial Bases}

\subsubsection{Monomial Polynomial} A non-orthogonal polynomial that uses powers of the input. It has the following form:
\begin{equation}
    \mathbf{P}_k=\hat{\textbf{A}}^k=\left(\mathbf{D}^{-\frac{1}{2}}\mathbf{A}\mathbf{D}^{-\frac{1}{2}}\right)^k
\end{equation}

\subsubsection{Chebyshev Polynomial} A widely used orthogonal polynomial. It is a special case of Jacobi polynomial when $a=b=-\frac{1}{2}$. It has the following form, when $k=0$ or $k=1$: 
\begin{equation}
\begin{split}
    \mathbf{T}_0(\hat{\textbf{A}})&=\mathbf{I} \\
    \mathbf{T}_1(\hat{\textbf{A}})&=\hat{\textbf{A}}
\end{split}
\end{equation}
\noindent when $k\ge 2$:
\begin{equation}
    \mathbf{T}_k=2\hat{\textbf{A}}\mathbf{T}_{k-1}(\hat{\textbf{A}})-\mathbf{T}_{k-2}(\hat{\textbf{A}})
\end{equation}

\subsubsection{Legendre Polynomial} Also a widely used orthogonal polynomial. It is a special case of Jacobi polynomial when $a=b=0$. It has the following form, when $k=0$ or $k=1$:
\begin{equation}
\begin{split}
    \mathbf{L}_0(\hat{\textbf{A}})&=\mathbf{I} \\
    \mathbf{L}_1(\hat{\textbf{A}})&=-\hat{\textbf{A}}+\textbf{I} \\
\end{split}
\end{equation}
\noindent when $k\ge 2$:
\begin{equation}
    k\mathbf{L}_k=(2k-1)\hat{\textbf{A}}\mathbf{L}_k(\hat{\textbf{A}})-(k-1)\mathbf{L}_{k-2}(\hat{\textbf{A}})
\end{equation}

\begin{figure*}[t]
    \centering
    \subfigure{\includegraphics[width=.23\linewidth]{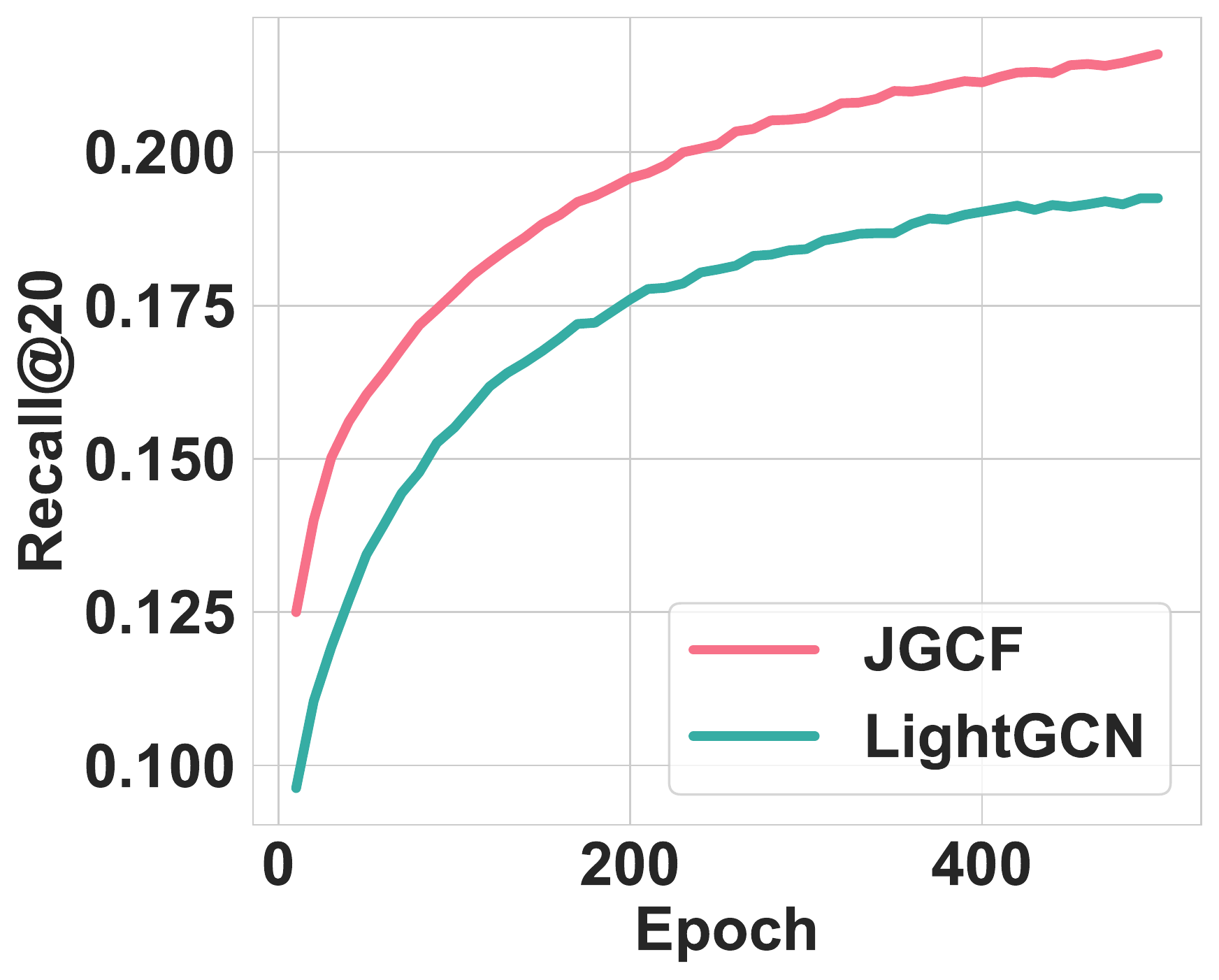}}
    \subfigure{\includegraphics[width=.23\linewidth]{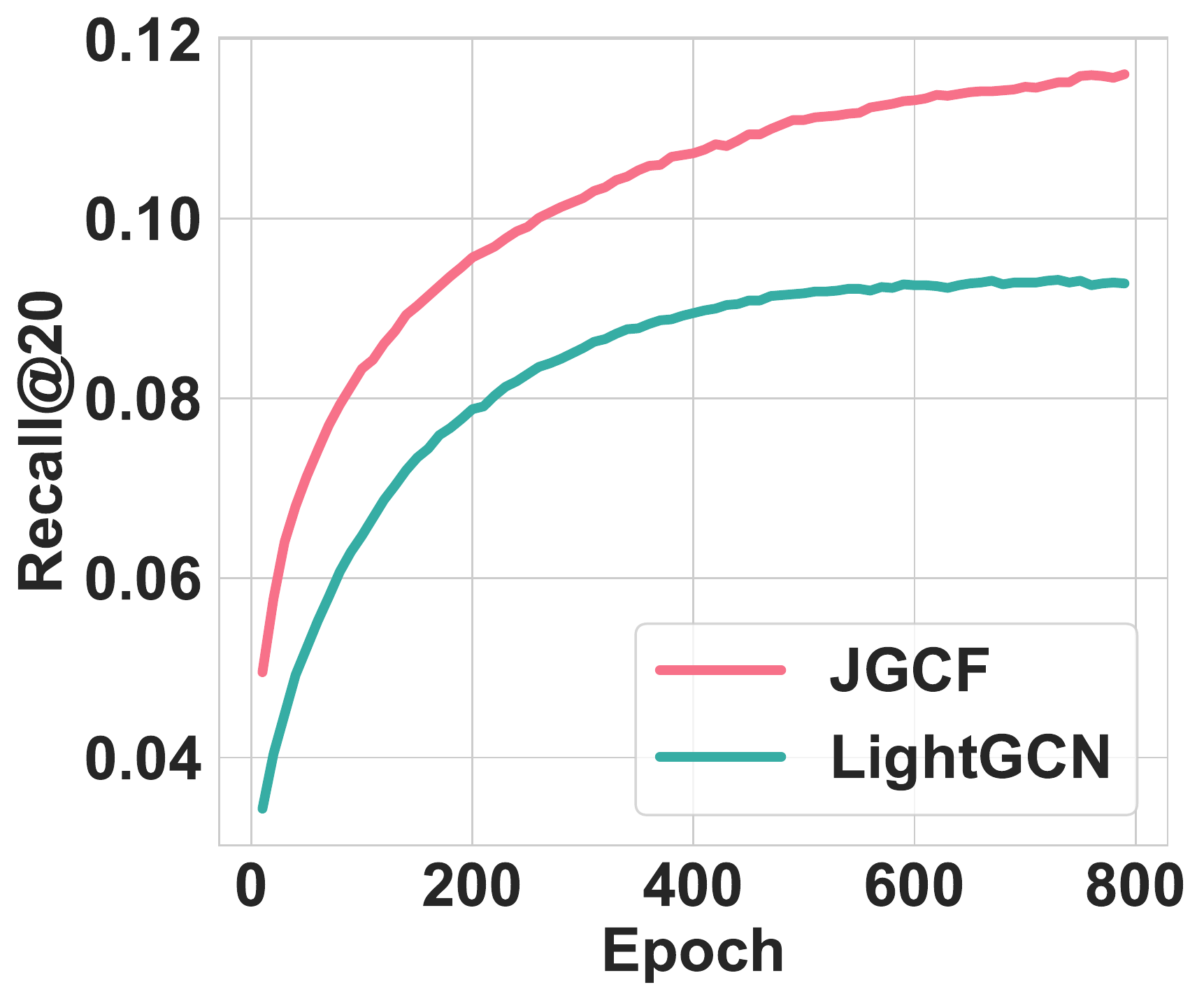}}
    \subfigure{\includegraphics[width=.23\linewidth]{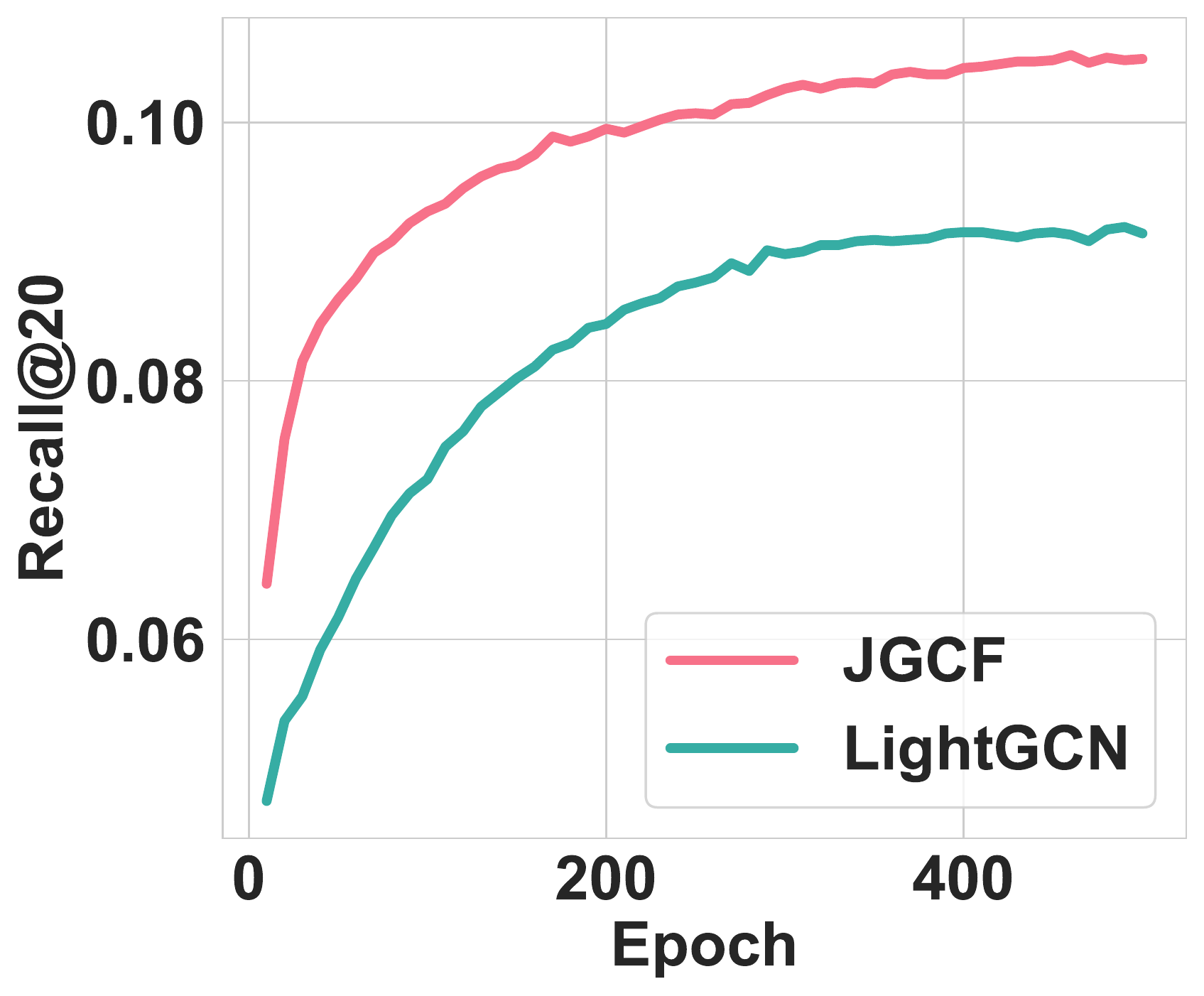}}
    \subfigure{\includegraphics[width=.23\linewidth]{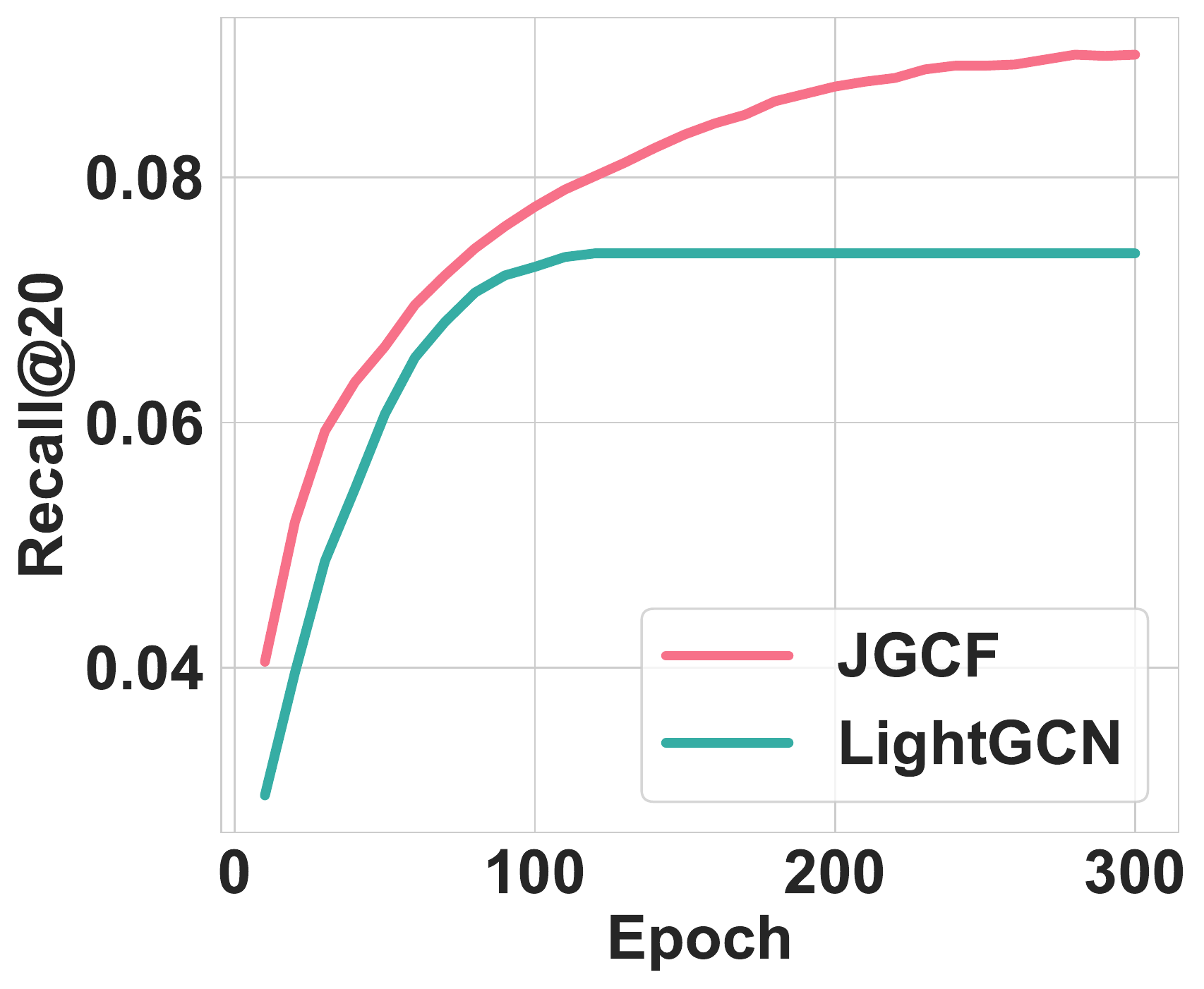}}
    \subfigure[Gowalla]{\includegraphics[width=.23\linewidth]{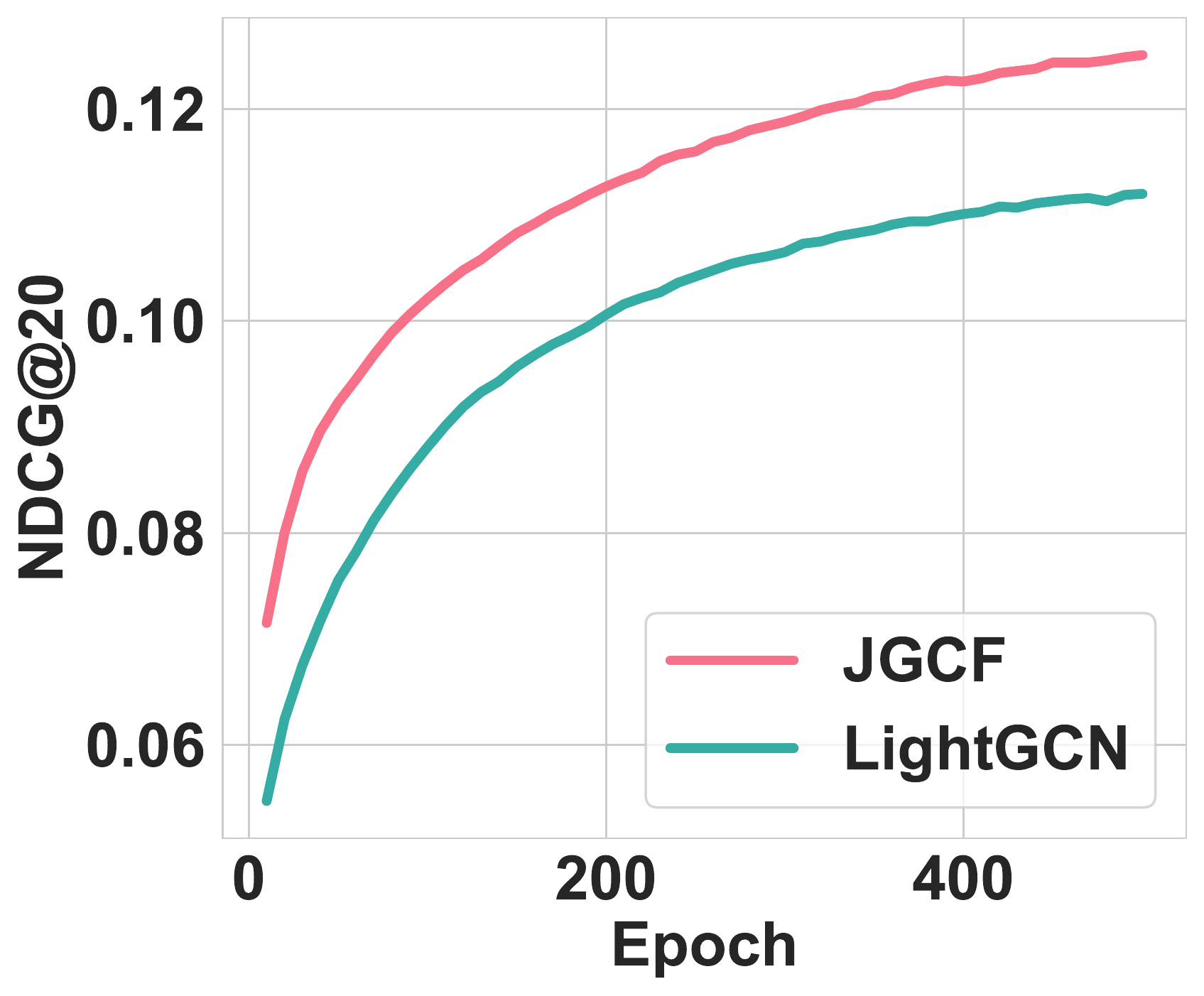}}
    \subfigure[Amazon-Books]{\includegraphics[width=.23\linewidth]{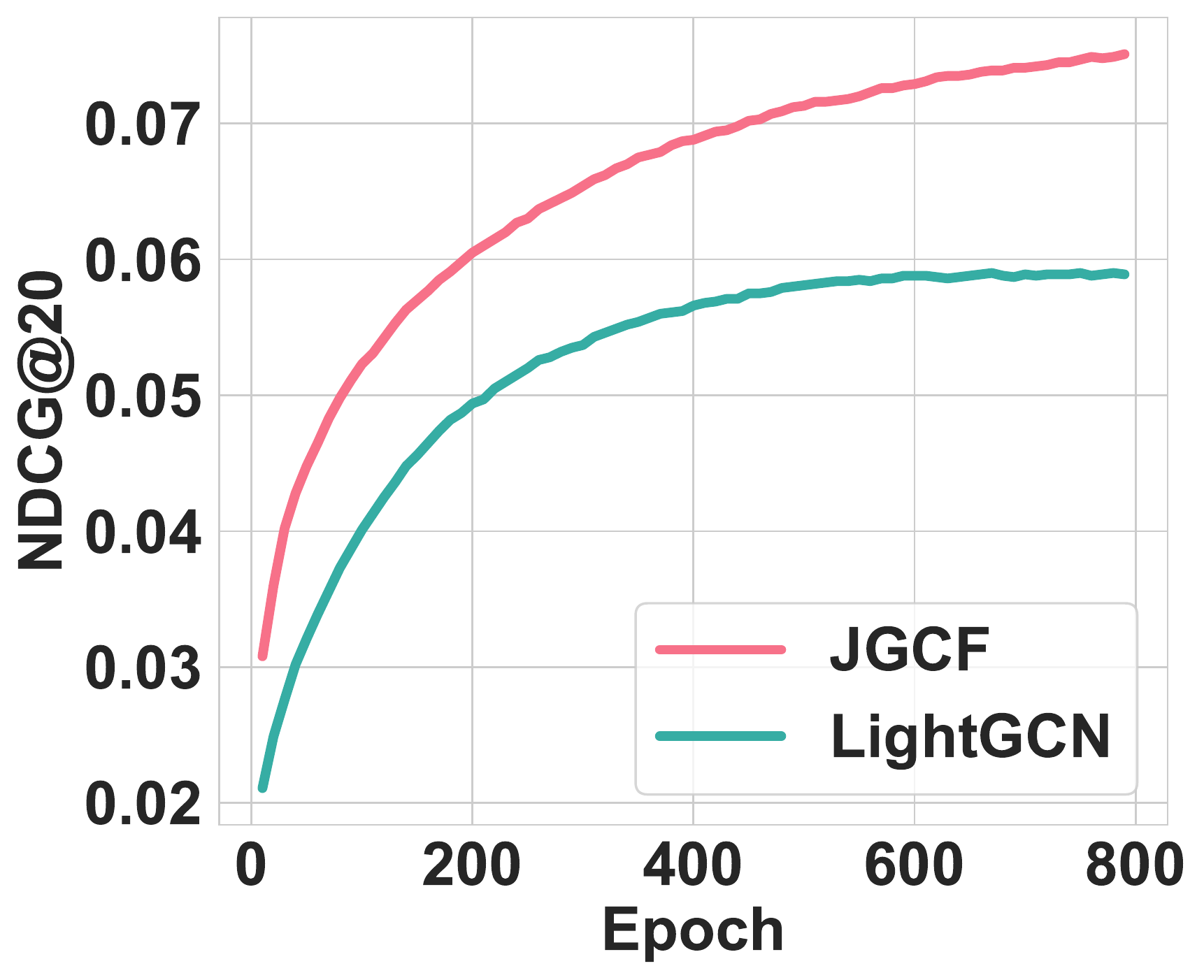}}
    \subfigure[Yelp]{\includegraphics[width=.23\linewidth]{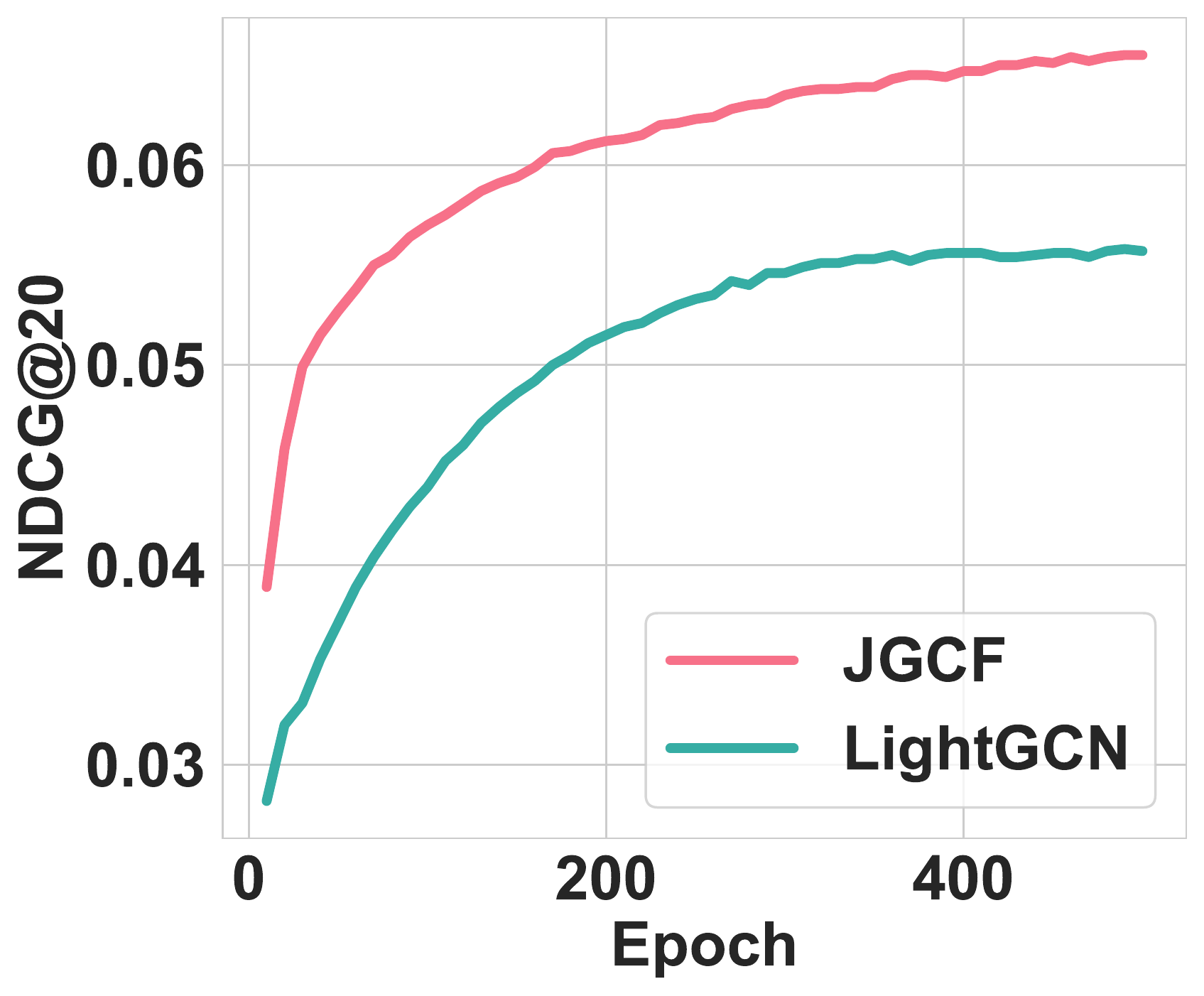}}
    \subfigure[Alibaba-iFashion]{\includegraphics[width=.23\linewidth]{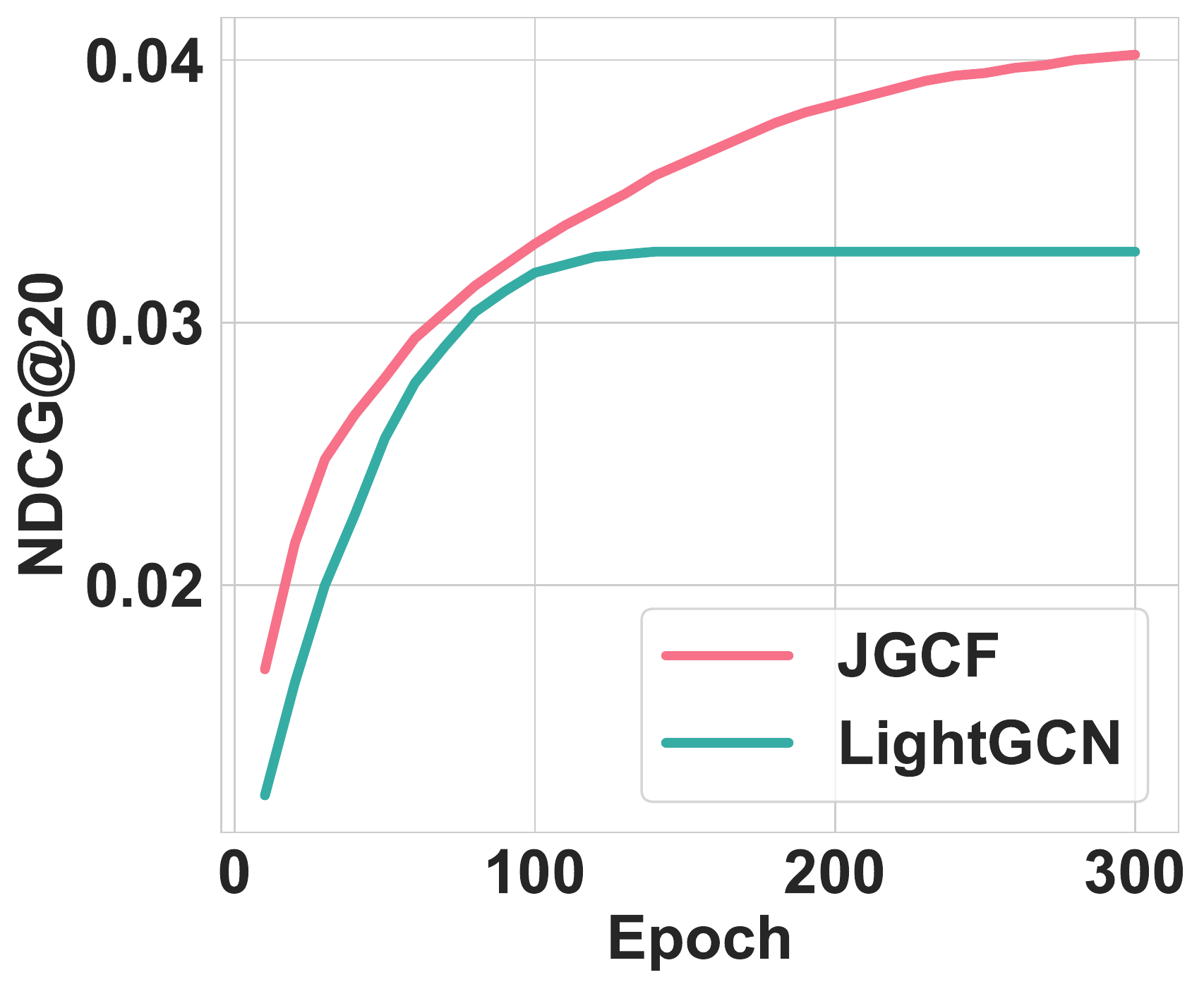}}
    \caption{Training Curves.}
    \label{fig:curve}
\end{figure*}

\subsection{Influence of $a$ and $b$}
\begin{figure}[t]
    \centering
    \subfigure{\includegraphics[width=.49\linewidth]{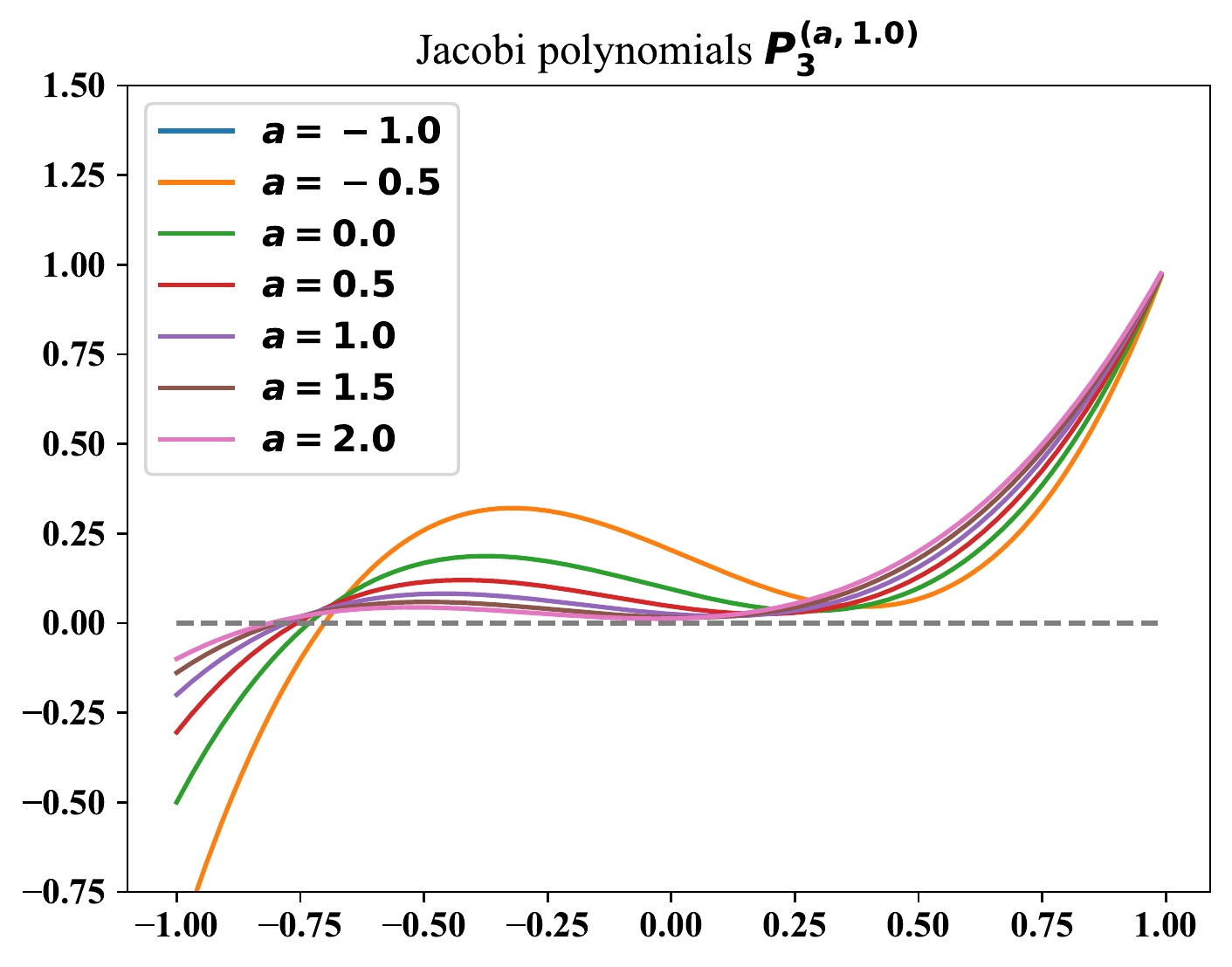}}
    \subfigure{\includegraphics[width=.48\linewidth]{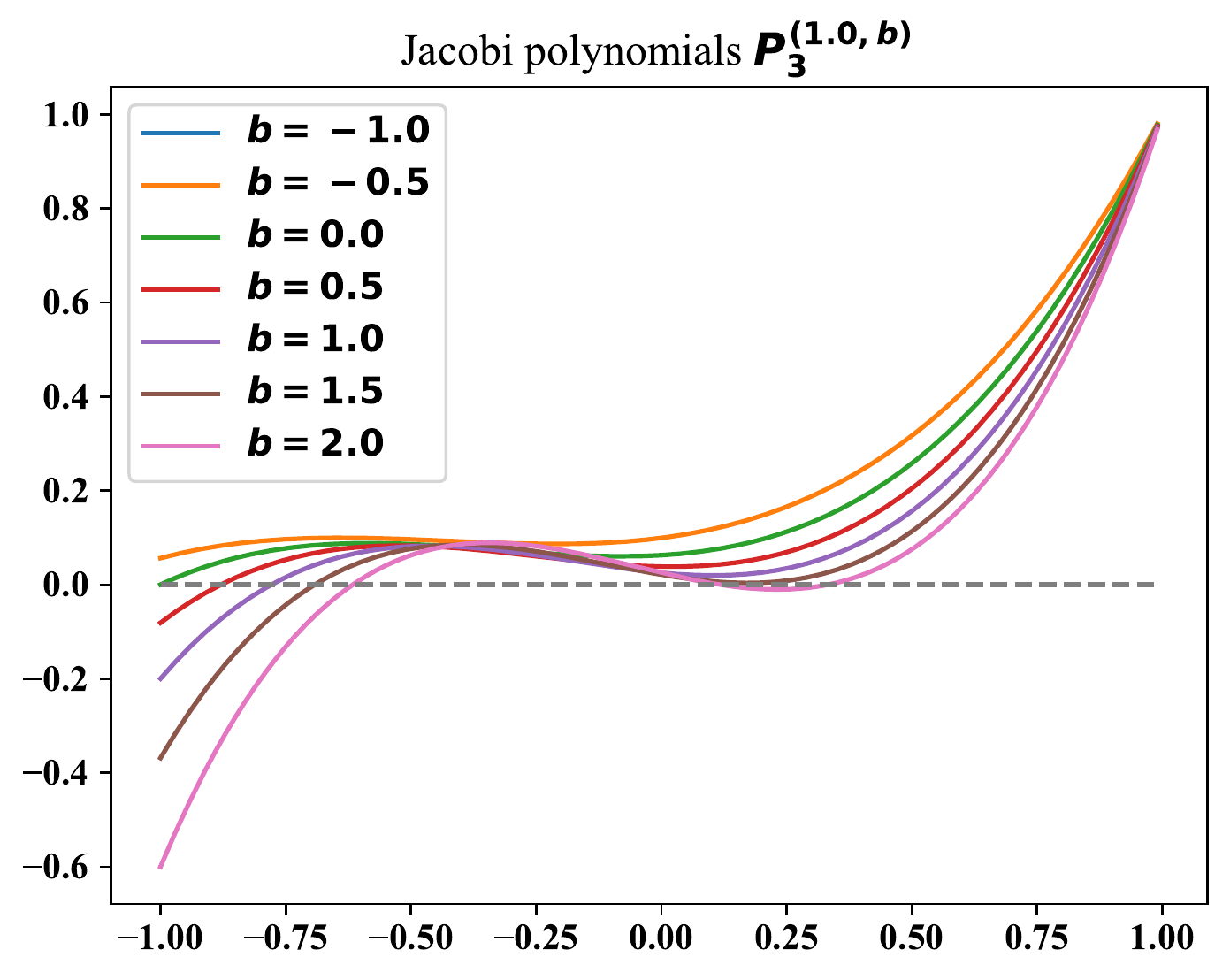}}
    \caption{Influence of $a$ and $b$ on Jacobi polynomials.}
    \label{fig:a_and_b}
\end{figure}

The influence of $a$ and $b$ are shown in Figure~\ref{fig:a_and_b}. We find when fixing $a$ increasing $b$ will allow more weight on the high-frequency signals. Besides, it will also suppress the low-frequency component.

\subsection{Training Curve}
To show the training of JGCF and LightGCN, we plot the Recall@20 and NDCG@20 on the validation set during training in Figure~\ref{fig:curve}. From the figure, we can observe that at the same epoch, JGCF always outperforms the other method and it is also not easy to overfit compared with LightGCN in most cases.

\subsection{Guidence for Choosing Hyper-parameter}

For very sparse graphs, we recommend increasing $b$ and $K$ while decreasing $a$ to suppress the middle-frequency signal further. While for denser graphs, we recommend using smaller $K$ and $b$ to weaken the suppression of middle-frequency signal for in denser graphs, the middle-frequency component may contain useful information that helps generalization too.

\end{document}